\documentclass[nohyperref]{article}

\usepackage{microtype}
\usepackage{graphicx}
\usepackage{subfigure}
\usepackage{booktabs} 
\usepackage[dvipsnames]{xcolor}
\usepackage{soul}
\usepackage{dblfloatfix}

\usepackage{hyperref}
\usepackage{multirow}

\usepackage[accepted]{icml2022}

\usepackage{amsmath}
\usepackage{amssymb}
\usepackage{mathtools}
\usepackage{amsthm}

\theoremstyle{plain}

\theoremstyle{definition}

\theoremstyle{remark}

\usepackage[textsize=tiny]{todonotes}

\icmltitlerunning{Robust Speech Recognition via Large-Scale Weak Supervision}
\begin{document}

\twocolumn[
\icmltitle{Robust Speech Recognition via Large-Scale Weak Supervision}



\icmlsetsymbol{equal}{*}

\begin{icmlauthorlist}
\icmlauthor{Alec Radford}{equal,openai}
\icmlauthor{Jong Wook Kim}{equal,openai}
\icmlauthor{Tao Xu}{openai}
\icmlauthor{Greg Brockman}{openai}
\icmlauthor{Christine McLeavey}{openai}
\icmlauthor{Ilya Sutskever}{openai}
\end{icmlauthorlist}

\icmlaffiliation{openai}{OpenAI, San Francisco, CA 94110, USA}

\icmlcorrespondingauthor{Alec Radford}{alec@openai.com}
\icmlcorrespondingauthor{Jong Wook Kim}{jongwook@openai.com}

\icmlkeywords{Machine Learning, ICML}

\vskip 0.3in
]



\printAffiliationsAndNotice{\icmlEqualContribution} 

\begin{abstract}
We study the capabilities of speech processing systems trained simply to predict large amounts of transcripts of audio on the internet. When scaled to 680,000 hours of multilingual and multitask supervision, the resulting models generalize well to standard benchmarks and are often competitive with prior fully supervised results but in a zero-shot transfer setting without the need for any fine-tuning. When compared to humans, the models approach their accuracy and robustness. We are releasing models and inference code to serve as a foundation for further work on robust speech processing.
\end{abstract}

\section{Introduction}\label{sec:introduction}

Progress in speech recognition has been energized by the development of unsupervised pre-training techniques exemplified by Wav2Vec 2.0 \cite{baevski2020wav2vec2}. Since these methods learn directly from raw audio without the need for human labels, they can productively use large datasets of unlabeled speech and have been quickly scaled up to 1,000,000 hours of training data \cite{zhang2021bigssl}, far more than the 1,000 or so hours typical of an academic supervised dataset. When fine-tuned on standard benchmarks, this approach has improved the state of the art, especially in a low-data setting.

These pre-trained audio encoders learn high-quality representations of speech, but because they are purely unsupervised they lack an equivalently performant decoder mapping those representations to usable outputs, necessitating a fine-tuning stage in order to actually perform a task such as speech recognition\footnote{\citet{baevski2021unsupervised} is an exciting exception - having developed a fully unsupervised speech recognition system}. This unfortunately limits their usefulness and impact as fine-tuning can still be a complex process requiring a skilled practitioner. There is an additional risk with requiring fine-tuning. Machine learning methods are exceedingly adept at finding patterns within a training dataset which boost performance on held-out data from the same dataset. However, some of these patterns are brittle and spurious and don't generalize to other datasets and distributions. In a particularly disturbing example, \citet{radford2021clip} documented a 9.2\% increase in object classification accuracy when fine-tuning a computer vision model on the ImageNet dataset \cite{russakovsky2015imagenet} without observing any improvement in average accuracy when classifying the same objects on seven other natural image datasets. A model that achieves ``superhuman'' performance when trained on a dataset can still make many basic errors when evaluated on another, possibly precisely because it is exploiting those dataset-specific quirks that humans are oblivious to \cite{geirhos2020shortcut}.

This suggests that while unsupervised pre-training has improved the quality of audio encoders dramatically, the lack of an equivalently high-quality pre-trained decoder, combined with a recommended protocol of dataset-specific fine-tuning, is a crucial weakness which limits their usefulness and robustness. The goal of a speech recognition system should be to work reliably ``out of the box'' in a broad range of environments without requiring supervised fine-tuning of a decoder for every deployment distribution.

As demonstrated by \citet{narayanan2018toward}, \citet{likhomanenko2020rethinking}, and \citet{chan2021speechstew} speech recognition systems that are pre-trained in a \textit{supervised} fashion across many datasets/domains exhibit higher robustness and generalize much more effectively to held-out datasets than models trained on a single source. These works achieve this by combining as many existing high-quality speech recognition datasets as possible. However, there is still only a moderate amount of this data easily available. SpeechStew \cite{chan2021speechstew} mixes together 7 pre-existing datasets totalling 5,140 hours of supervision. While not insignificant, this is still tiny compared to the previously mentioned 1,000,000 hours of unlabeled speech data utilized in \citet{zhang2021bigssl}.

Recognizing the limiting size of existing high-quality supervised datasets, recent efforts have created larger datasets for speech recognition. By relaxing the requirement of gold-standard human-validated transcripts, \citet{chen2021gigaspeech} and \citet{galvez2021people} make use of sophisticated automated pipelines to scale weakly supervised speech recognition to 10,000 and 30,000 hours of noisier training data. This trade-off between quality and quantity is often the right call. Although understudied so far for speech recognition, recent work in computer vision has demonstrated that moving beyond gold-standard crowdsourced datasets such as ImageNet \cite{russakovsky2015imagenet} to much larger but weakly supervised datasets significantly improves the robustness and generalization of models \cite{mahajan2018exploring,kolesnikov2020big}.

Yet these new datasets are only a few times larger than the sum of existing high-quality datasets and still much smaller than prior unsupervised work. In this work we close that gap, scaling weakly supervised speech recognition the next order of magnitude to 680,000 hours of labeled audio data. We call our approach Whisper\footnote{If an acronym or basis for the name is desired, WSPSR standing for \textbf{W}eb-scale \textbf{S}upervised \textbf{P}retraining for \textbf{S}peech \textbf{R}ecognition can be used.}. We demonstrate models trained at this scale transfer well to existing datasets zero-shot, removing the need for any dataset-specific fine-tuning to achieve high-quality results.

In addition to scale, our work also focuses on broadening the scope of weakly supervised pre-training beyond English-only speech recognition to be both multilingual and multitask. Of those 680,000 hours of audio, 117,000 hours cover 96 other languages. The dataset also includes 125,000 hours of \texttt{X$\rightarrow$en} translation data. We find that for sufficiently large models there is no drawback and even benefits to joint multilingual and multitask training.

Our work suggests that simple scaling of weakly supervised pre-training has been underappreciated so far for speech recognition. We achieve these results without the need for the self-supervision or self-training techniques that have been a mainstay of recent large-scale speech recognition work. To serve as a foundation for further research on robust speech recognition, we release inference code and models at the following URL: \url{https://github.com/openai/whisper}.

\section{Approach}\label{sec:approach}

\subsection{Data Processing}\label{subsec:data}

Following the trend of recent work leveraging web-scale text from the internet for training machine learning systems, we take a minimalist approach to data pre-processing. In contrast to a lot of work on speech recognition, we train Whisper models to predict the raw text of transcripts without any significant standardization, relying on the expressiveness of sequence-to-sequence models to learn to map between utterances and their transcribed form. This simplifies the speech recognition pipeline since it removes the need for a separate inverse text normalization step in order to produce naturalistic transcriptions. 

We construct the dataset from audio that is paired with transcripts on the Internet. This results in a very diverse dataset covering a broad distribution of audio from many different environments, recording setups, speakers, and languages. While diversity in audio quality can help train a model to be robust, diversity in transcript quality is not similarly beneficial. Initial inspection showed a large amount of subpar transcripts in the raw dataset. To address this, we developed several automated filtering methods to improve transcript quality. 

Many transcripts on the internet are not actually human-generated but the output of existing ASR systems. Recent research has shown that training on datasets of mixed human and machine-generated data can significantly impair the performance of translation systems \cite{ghorbani2021scaling}. In order to avoid learning ``transcript-ese'', we developed many heuristics to detect and remove machine-generated transcripts from the training dataset. Many existing ASR systems output only a limited subset of written language which removes or normalizes away aspects that are difficult to predict from only audio signals such as complex punctuation (exclamation points, commas, and question marks), formatting whitespace such as paragraphs, or stylistic aspects such as capitalization. An all-uppercase or all-lowercase transcript is very unlikely to be human generated. While many ASR systems include some level of inverse text normalization, it is often simple or rule-based and still detectable from other unhandled aspects such as never including commas.

We also use an audio language detector, which was created by fine-tuning a prototype model trained on a prototype version of the dataset on VoxLingua107 \cite{valk2021voxlingua107} to ensure that the spoken language matches the language of the transcript according to CLD2. If the two do not match, we don't include the (audio, transcript) pair as a speech recognition training example in the dataset. We make an exception if the transcript language is English and add these pairs to the dataset as \texttt{X$\rightarrow$en} speech translation training examples instead. We use fuzzy de-duping of transcript texts to reduce the amount of duplication and automatically generated content in the training dataset.

We break audio files into 30-second segments paired with the subset of the transcript that occurs within that time segment. We train on all audio, including segments where there is no speech (though with sub-sampled probability) and use these segments as training data for voice activity detection.

For an additional filtering pass, after training an initial model we aggregated information about its error rate on training data sources and performed manual inspection of these data sources sorting by a combination of both high error rate and data source size in order to identify and remove low-quality ones efficiently. This inspection showed a large amount of only partially transcribed or poorly aligned/misaligned transcripts as well as remaining low-quality machine-generated captions that filtering heuristics did not detect.

To avoid contamination, we perform de-duplication at a transcript level between the training dataset and the evaluation datasets we thought were at higher risk of overlap, namely TED-LIUM 3 \cite{Hernandez2018TEDLIUM3T}.

\begin{figure*}[t]
\begin{center}
\centerline{\includegraphics[width=1.0\textwidth]{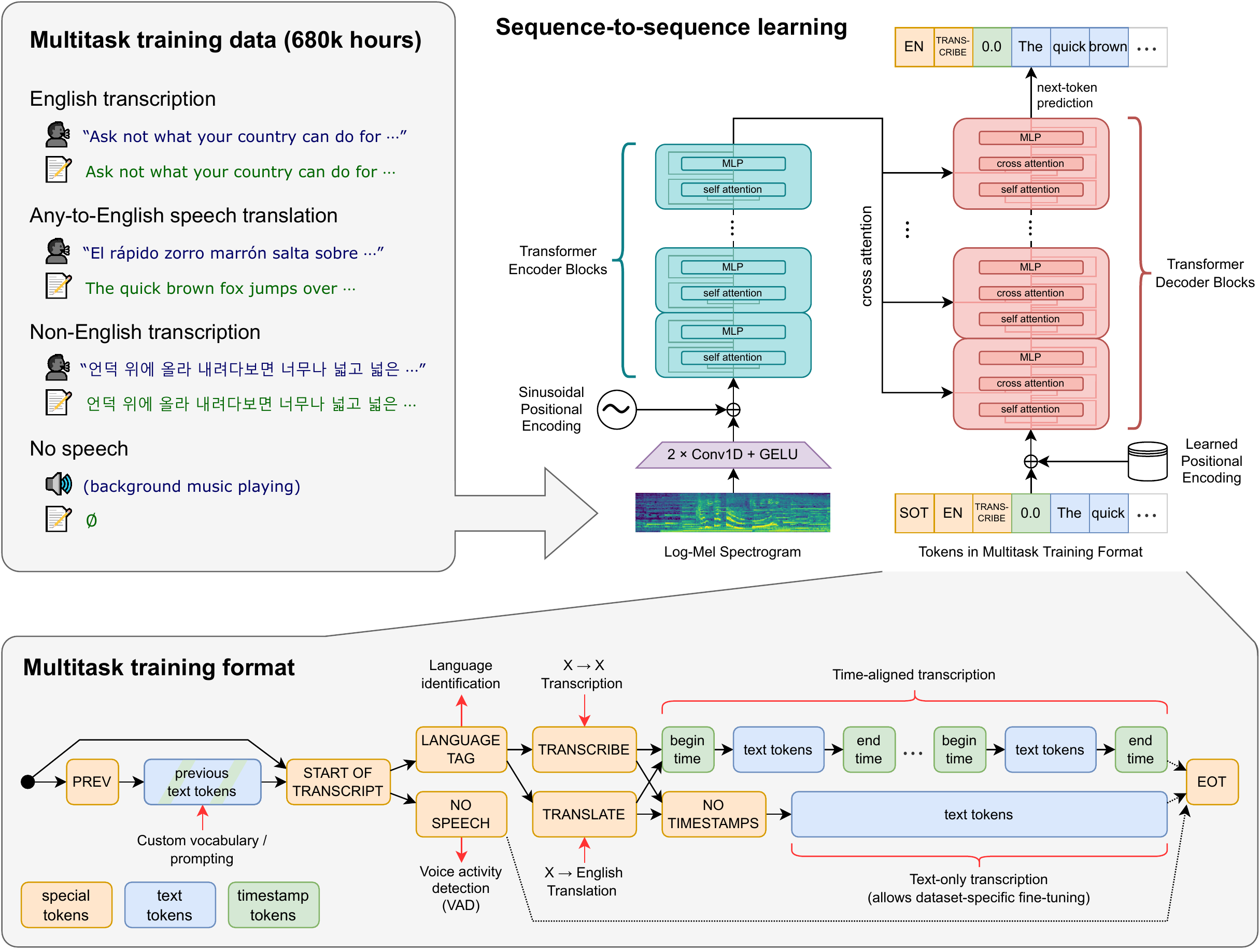}}
\caption{\textbf{Overview of our approach.} A sequence-to-sequence Transformer model is trained on many different speech processing tasks, including multilingual speech recognition, speech translation, spoken language identification, and voice activity detection. All of these tasks are jointly represented as a sequence of tokens to be predicted by the decoder, allowing for a single model to replace many different stages of a traditional speech processing pipeline. The multitask training format uses a set of special tokens that serve as task specifiers or classification targets, as further explained in Section \ref{sec:multitask}.}
\label{fig:approach}
\end{center}
\vspace{-1em}
\end{figure*}

\subsection{Model}\label{subsec:model}

Since the focus of our work is on studying the capabilities of large-scale supervised pre-training for speech recognition, we use an off-the-shelf architecture to avoid confounding our findings with model improvements. We chose an encoder-decoder Transformer \cite{vaswani2017transformer} as this architecture has been well validated to scale reliably. All audio is re-sampled to 16,000 Hz, and an 80-channel log-magnitude Mel spectrogram representation is computed on 25-millisecond windows with a stride of 10 milliseconds. For feature normalization, we globally scale the input to be between -1 and 1 with approximately zero mean across the pre-training dataset. The encoder processes this input representation with a small stem consisting of two convolution layers with a filter width of 3 and the GELU activation function \cite{hendrycks2016gaussian} where the second convolution layer has a stride of two. Sinusoidal position embeddings are then added to the output of the stem after which the encoder Transformer blocks are applied. The transformer uses pre-activation residual blocks \cite{child2019generating}, and a final layer normalization is applied to the encoder output. The decoder uses learned position embeddings and tied input-output token representations \cite{press-wolf-2017-using}. The encoder and decoder have the same width and number of transformer blocks.
Figure \ref{fig:approach} summarizes the model architecture.

We use the same byte-level BPE text tokenizer used in GPT-2 \cite{sennrich2015neural,radford2019gpt2} for the English-only models and refit the vocabulary (but keep the same size) for the multilingual models to avoid excessive fragmentation on other languages since the GPT-2 BPE vocabulary is English only.

\subsection{Multitask Format}\label{sec:multitask}

Although predicting which words were spoken in a given audio snippet is a core part of the full speech recognition problem and extensively studied in research, it is not the only part. A fully featured speech recognition system can involve many additional components such as voice activity detection, speaker diarization, and inverse text normalization. These components are often handled separately, resulting in a relatively complex system around the core speech recognition model. To reduce this complexity, we would like to have a single model perform the entire speech processing pipeline, not just the core recognition part. An important consideration here is the interface for the model. There are many different tasks that can be performed on the same input audio signal: transcription, translation, voice activity detection, alignment, and language identification are some examples. 

For this kind of one-to-many mapping to work with a single model, some form of task specification is necessary. We use a simple format to specify all tasks and conditioning information as a sequence of input tokens to the decoder. Since our decoder is an audio-conditional language model, we also train it to condition on the history of text of the transcript in the hope that it will learn to use longer-range text context to resolve ambiguous audio. Specifically, with some probability we add the transcript text preceding the current audio segment to the decoder's context. We indicate the beginning of prediction with a \texttt{<|startoftranscript|>} token. First, we predict the language being spoken which is represented by a unique token for each language in our training set (99 total). These language targets are sourced from the aforementioned VoxLingua107 model. In the case where there is no speech in an audio segment, the model is trained to predict a \texttt{<|nospeech|>} token indicating this. The next token specifies the task (either transcription or translation) with an \texttt{<|transcribe|>} or \texttt{<|translate|>} token. After this, we specify whether to predict timestamps or not by including a \texttt{<|notimestamps|>} token for that case. At this point, the task and desired format is fully specified, and the output begins. For timestamp prediction, we predict time relative to the current audio segment, quantizing all times to the nearest 20 milliseconds which matches the native time resolution of Whisper models, and add additional tokens to our vocabulary for each of these. We interleave their prediction with the caption tokens: the start time token is predicted before each caption's text, and the end time token is predicted after. When a final transcript segment is only partially included in the current 30-second audio chunk, we predict only its start time token for the segment when in timestamp mode, to indicate that the subsequent decoding should be performed on an audio window aligned with that time, otherwise we truncate the audio to not include the segment. Lastly, we add a \texttt{<|endoftranscript|>} token. We only mask out the training loss over the previous context text, and train the model to predict all other tokens. Please see Figure \ref{fig:approach} for an overview of our format and training setup.

\subsection{Training Details}\label{subsec:training-details}

We train a suite of models of various sizes in order to study the scaling properties of Whisper. Please see Table \ref{tab:models} for an overview. We train with data parallelism across accelerators using FP16 with dynamic loss scaling and activation checkpointing \cite{griewank2000algorithm, chen2016training}. Models were trained with AdamW \cite{loshchilov2017decoupled} and gradient norm clipping \cite{pascanu2013difficulty} with a linear learning rate decay to zero after a warmup over the first 2048 updates. A batch size of 256 segments was used, and the models are trained for 2\textsuperscript{20} updates which is between two and three passes over the dataset. Due to only training for a few epochs, over-fitting is not a large concern, and we do not use any data augmentation or regularization and instead rely on the diversity contained within such a large dataset to encourage generalization and robustness. Please see Appendix \ref{sec:hyperparameters} for full training hyperparameters.\footnote{After the original release of Whisper, we trained an additional Large model (denoted V2) for 2.5X more epochs while adding SpecAugment \cite{park2019specaugment}, Stochastic Depth \cite{huang2016deep}, and BPE Dropout \cite{provilkov2019bpe} for regularization. Reported results have been updated to this improved model unless otherwise specified.}

During early development and evaluation we observed that Whisper models had a tendency to transcribe plausible but almost always incorrect guesses for the names of speakers. This happens because many transcripts in the pre-training dataset include the name of the person who is speaking, encouraging the model to try to predict them, but this information is only rarely inferable from only the most recent 30 seconds of audio context. To avoid this, we fine-tune Whisper models briefly on the subset of transcripts that do not include speaker annotations which removes this behavior.

\begin{table}[t]
\centering
\small
\begin{tabular}{l|cccc} \toprule
    Model & Layers & Width & Heads & Parameters \\
    \midrule
    Tiny & 4 & 384 & 6 & 39M \\
    Base & 6 & 512 & 8 & 74M \\
    Small & 12 & 768 & 12 & 244M \\
    Medium & 24 & 1024 & 16 & 769M \\
    Large & 32 & 1280 & 20 & 1550M \\
    \bottomrule
\end{tabular}
\caption{Architecture details of the Whisper model family.}\label{tab:models}
\end{table}

\section{Experiments}\label{sec:experiments}

\subsection{Zero-shot Evaluation}\label{subsec:zeroshot}

The goal of Whisper is to develop a single robust speech processing system that works reliably without the need for dataset specific fine-tuning to achieve high-quality results on specific distributions. To study this capability, we re-use a wide set of existing speech processing datasets to check whether Whisper is able to generalize well across domains, tasks, and languages. Instead of using the standard evaluation protocol for these datasets, which include both a train and test split, we evaluate Whisper in a zero-shot setting without using any of the training data for each of these datasets so that we are measuring broad generalization.

\subsection{Evaluation Metrics}\label{subsec:eval-metric}

Speech recognition research typically evaluates and compares systems based on the word error rate (WER) metric. However, WER, which is based on string edit distance, penalizes all differences between the model's output and the reference transcript including innocuous differences in transcript style. As a result, systems that output transcripts that would be judged as correct by humans can still have a large WER due to minor formatting differences. While this poses a problem for all transcribers, it is particularly acute for zero-shot models like Whisper, which do not observe any examples of specific datasets transcript formats.

This is not a novel observation; the development of evaluation metrics that better correlate with human judgement is an active area of research, and while there are some promising methods, none have seen widespread adoption for speech recognition yet. We opt to address this problem with extensive standardization of text before the WER calculation to minimize penalization of non-semantic differences. Our text normalizer was developed through iterative manual inspection to identify common patterns where naive WER penalized Whisper models for an innocuous difference. Appendix \ref{sec:standardization} includes full details. For several datasets, we observe WER drops of up to 50 percent usually due to a quirk such as a dataset's reference transcripts seperating contractions from words with whitespace. We caution this development procedure comes at a risk of overfitting to the transcription style of Whisper models which we investigate in Section \ref{subsec:text-normalization-analysis}. We are releasing the code for our text normalizer to allow for easy comparison and to help others study the performance of speech recognition systems in out-of-distribution settings.

\subsection{English Speech Recognition}\label{subsec:zero-shot}

In 2015, Deep Speech 2 \cite{amodei2015deepspeech2} reported a speech recognition system matched human-level performance when transcribing the LibriSpeech test-clean split. As part of their analysis they concluded: \textit{``Given this result, we suspect that there is little room for a generic speech system to further improve on clean read speech without further domain adaptation.''} Yet seven years later the SOTA WER on LibriSpeech test-clean has dropped another 73\% from their 5.3\% to 1.4\% \cite{zhang2021bigssl}, far below their reported human-level error rate of 5.8\%. Despite this massive and unanticipated further improvement in performance on held-out but in-distribution data, speech recognition models trained on LibriSpeech remain far above human error rates when used in other settings.
What explains this gap between reportedly superhuman performance in-distribution and subhuman performance out-of-distribution?

We suspect a large part of this gap between human and machine behavior is due to conflating different capabilities being measured by human and machine performance on a test set. This claim may seem confusing at first; if both humans and machines are taking the same test, how can it be that different skills are being tested? The difference arises not in the testing but in how they trained for it. Humans are often asked to perform a task given little to no supervision on the specific data distribution being studied. Thus human performance is a measure of out-of-distribution generalization. But machine learning models are usually evaluated after training on a large amount of supervision from the evaluation distribution, meaning that machine performance is instead a measure of in-distribution generalization. While both humans and machines are being evaluated on the same \textit{test} data, two quite different abilities are being measured due to a difference in \textit{train} data.

Whisper models, which are trained on a broad and diverse distribution of audio and evaluated in a zero-shot setting, could potentially match human behavior much better than existing systems. To study whether this is the case (or whether the difference between machine and human performance is due to yet-to-be-understood factors) we can compare Whisper models with both human performance and standard fine-tuned machine learning models and check which they more closely match.

\definecolor{Highlight}{HTML}{39b54a}  

\begin{figure}[t]
\begin{center}
\centerline{\includegraphics[width=1.0\columnwidth]{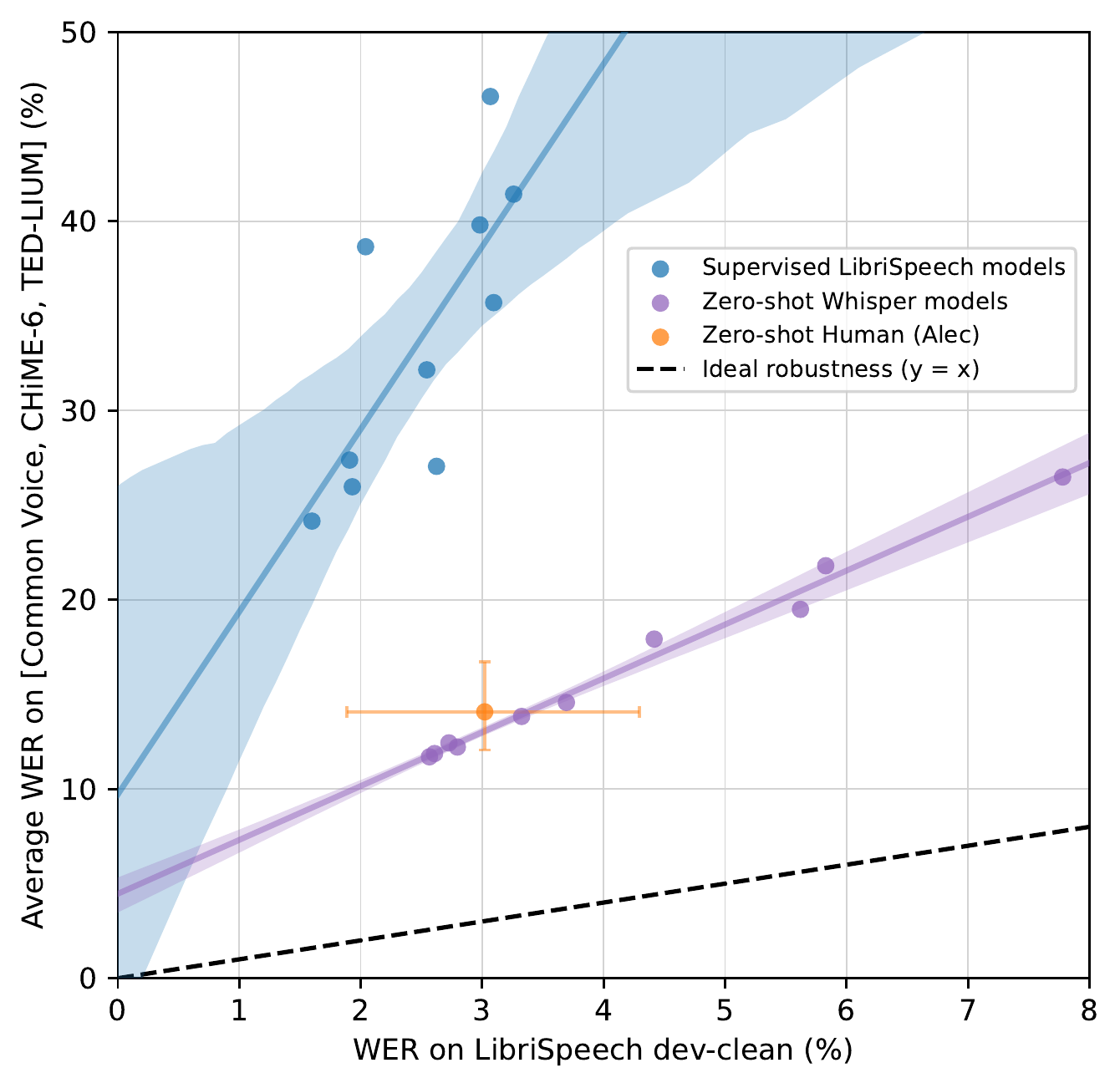}}
\caption{\textbf{Zero-shot Whisper models close the gap to human robustness.} Despite matching or outperforming a human on LibriSpeech dev-clean, supervised LibriSpeech models make roughly twice as many errors as a human on other datasets demonstrating their brittleness and lack of robustness. The estimated robustness frontier of zero-shot Whisper models, however, includes the 95\% confidence interval for this particular human.}
\label{robustness_figure}
\end{center}
\vspace{-1em}
\end{figure}

\begin{table}[t]
\vskip 0.15in
\small
\begin{center}
\begin{tabular}{l|cc|c}
\toprule
& wav2vec 2.0 & Whisper & RER \\
Dataset & Large (no LM) & Large V2 & (\%) \\
\midrule
LibriSpeech Clean & \textbf{2.7} & \textbf{2.7} & 0.0 \\
\midrule
Artie & 24.5 & \textbf{6.2} & \color{Highlight}74.7 \\
Common Voice & 29.9 & \textbf{9.0} & \color{Highlight}69.9 \\
Fleurs En & 14.6 & \textbf{4.4} & \color{Highlight}69.9 \\
Tedlium & 10.5 & \textbf{4.0} & \color{Highlight}61.9 \\
CHiME6 & 65.8 & \textbf{25.5} & \color{Highlight}61.2 \\
VoxPopuli En & 17.9 & \textbf{7.3} & \color{Highlight}59.2 \\
CORAAL & 35.6 & \textbf{16.2} & \color{Highlight}54.5 \\
AMI IHM & 37.0 & \textbf{16.9} & \color{Highlight}54.3 \\
Switchboard & 28.3 & \textbf{13.8} & \color{Highlight}51.2 \\
CallHome & 34.8 & \textbf{17.6} & \color{Highlight}49.4 \\
WSJ & 7.7 & \textbf{3.9} & \color{Highlight}49.4 \\
AMI SDM1 & 67.6 & \textbf{36.4} & \color{Highlight}46.2 \\
LibriSpeech Other & 6.2 & \textbf{5.2} & \color{Highlight}16.1 \\
\midrule
Average & 29.3 & \textbf{12.8} & \color{Highlight}55.2 \\

\bottomrule
\end{tabular}
\caption{\textbf{Detailed comparison of effective robustness across various datasets.} Although both models perform within 0.1\% of each other on LibriSpeech, a zero-shot Whisper model performs much better on other datasets than expected for its LibriSpeech performance and makes 55.2\% less errors on average. Results reported in word error rate (WER) for both models after applying our text normalizer.}
\label{robustness_table}
\end{center}
\vspace{-1em}
\end{table}

To quantify this difference, we examine both \textit{overall} robustness, that is average performance across many distributions/datasets, and \textit{effective} robustness, introduced by \citet{taori2020robustness}, which measures the difference in expected performance between a reference dataset, which is usually in-distribution, and one or more out-of-distribution datasets. A model with high effective robustness does better than expected on out-of-distribution datasets as a function of its performance on the reference dataset and approaches the ideal of equal performance on all datasets. For our analysis, we use LibriSpeech as the reference dataset due to its central role in modern speech recognition research and the availability of many released models trained on it, which allows for characterizing robustness behaviors. We use a suite of 12 other academic speech recognition datasets to study out-of-distribution behaviors. Full details about these datasets can be found in Appendix \ref{sec:datasets}.

Our main findings are summarized in Figure \ref{robustness_figure} and Table \ref{robustness_table}. Although the best zero-shot Whisper model has a relatively unremarkable LibriSpeech clean-test WER of 2.5, which is roughly the performance of modern supervised baseline or the mid-2019 state of the art, zero-shot Whisper models have very different robustness properties than supervised LibriSpeech models and out-perform all benchmarked LibriSpeech models by large amounts on other datasets. Even the smallest zero-shot Whisper model, which has only 39 million parameters and a 6.7 WER on LibriSpeech test-clean is roughly competitive with the best supervised LibriSpeech model when evaluated on other datasets. When compared to a human in Figure \ref{robustness_figure}, the best zero-shot Whisper models roughly match their accuracy and robustness. For a detailed breakdown of this large improvement in robustness, Table \ref{robustness_table} compares the performance of the best zero-shot Whisper model with a supervised LibriSpeech model that has the closest performance to it on LibriSpeech test-clean. Despite their very close performance on the reference distribution, the zero-shot Whisper model achieves an average relative error reduction of 55.2\% when evaluated on other speech recognition datasets. 

This finding suggests emphasizing zero-shot and out-of-distribution evaluations of models, particularly when attempting to compare to human performance, to avoid over-stating the capabilities of machine learning systems due to misleading comparisons.

\subsection{Multi-lingual Speech Recognition}\label{subsec:multi-lingual}

\begin{figure}[t]
\begin{center}
\centerline{\includegraphics[width=1.0\columnwidth]{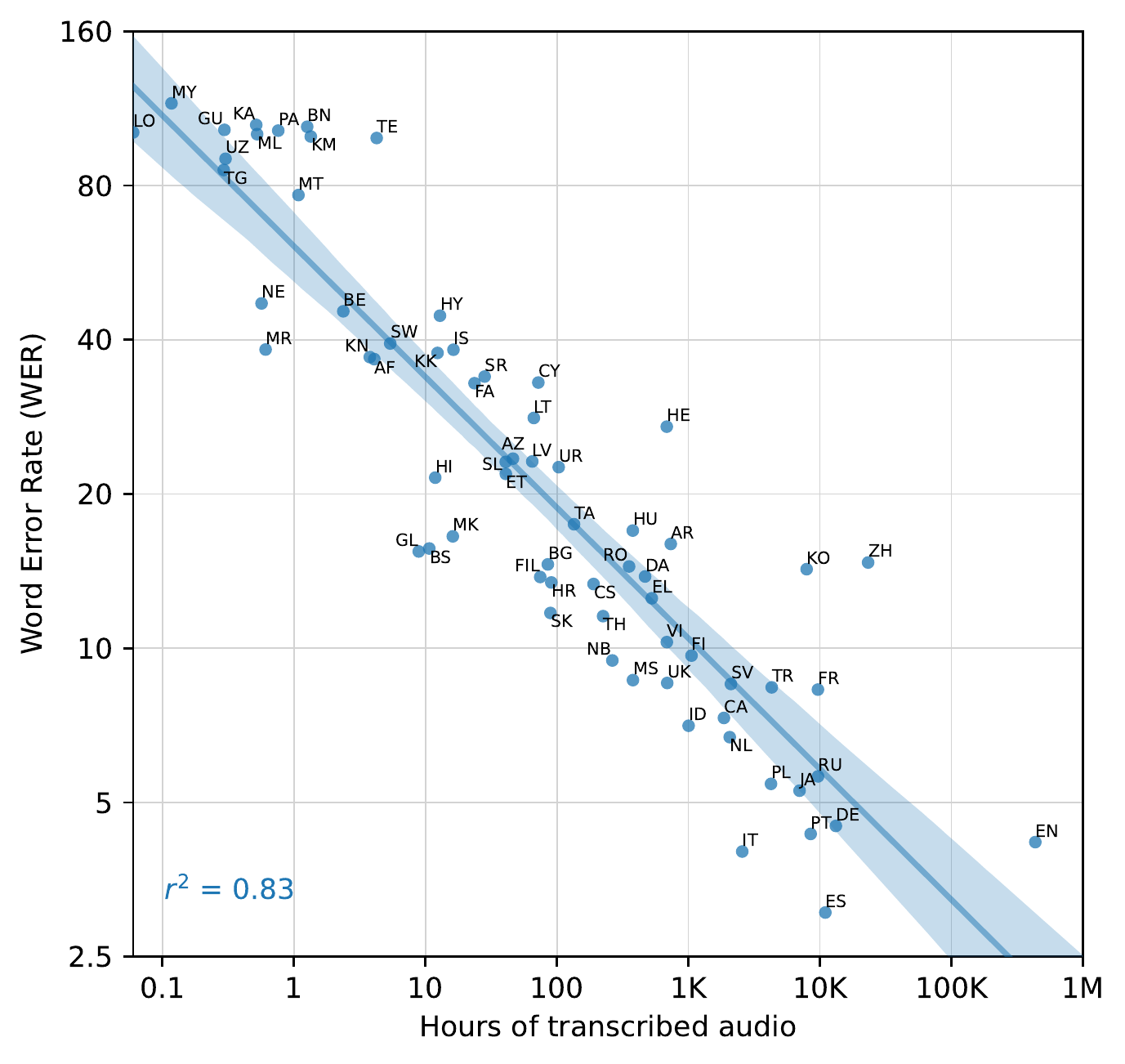}}
\caption{\textbf{Correlation of pre-training supervision amount with downstream speech recognition performance.} The amount of pre-training speech recognition data for a given language is very predictive of zero-shot performance on that language in Fleurs.}
\label{fleurs_asr_transfer}
\end{center}
\vspace{-1em}
\end{figure}

\begin{table}[t]
\vskip 0.15in
\begin{center}
\begin{tabular}{l|cc}
\toprule
Model & MLS & VoxPopuli \\
\midrule
VP-10K + FT & - & 15.3 \\
XLS-R (1B) & 10.9 & 10.6 \\
mSLAM-CTC (2B) & 9.7 & 9.1 \\
Maestro & - & \textbf{8.1} \\
\midrule
Zero-Shot Whisper & \textbf{7.3} & 13.6 \\
\bottomrule
\end{tabular}
\caption{\textbf{Multilingual speech recognition performance.} Zero-shot Whisper improves performance on Multilingual LibriSpeech (MLS) but is still significantly behind both Maestro, XLS-R, and mSLAM on VoxPopuli.}
\label{multilingual_table}
\end{center}
\vspace{-1em}
\end{table}

In order to compare to prior work on multilingual speech recognition, we report results on two low-data benchmarks: Multilingual LibriSpeech (MLS) \cite{pratap2020mls} and VoxPopuli \cite{wang2021voxpopuli} in Table \ref{multilingual_table}.

Whisper performs well on Multilingual LibriSpeech, outperforming XLS-R \cite{babu2021xlsr}, mSLAM \cite{bapna2022mslam}, and Maestro \cite{chen2022maestro} in a zero-shot setting. We caution that we do use a simple text standardizer for this result which prevents direct comparison or claims of SOTA performance. On VoxPopuli, however, Whisper significantly underperforms prior work and only beats the VP-10K+FT baseline from the original paper. We suspect the underperformance of Whisper models on VoxPopuli could be due to other models including this distribution as a major source for their unsupervised pre-training data and the dataset having significantly more supervised data, which benefits fine-tuning. While MLS has 10 hours of training data per language, the average amount of training data per language is roughly 10$\times$ higher for VoxPopuli.

These two benchmarks are somewhat narrow since they only include 15 unique languages, almost all of which are in the Indo-European language family and many of which are high-resource languages. These benchmarks only provide limited coverage and room to study Whisper models multilingual capabilities which include training data for speech recognition in 75 languages. To study the performance of Whisper more broadly we also report performance on the Fleurs dataset \cite{conneau2022fleurs}. In particular, we were interested in studying the relationship between the amount of training data we have for a given language and the resulting downstream zero-shot performance for that language. We visualize this relation in Figure \ref{fleurs_asr_transfer}. We find a strong squared correlation coefficient of 0.83 between the log of the word error rate and the log of the amount of training data per language. Checking the regression coefficient for a linear fit to these log-log values results in an estimate that WER halves for every 16$\times$ increase in training data. We also observed that many of the largest outliers in terms of worse than expected performance according to this trend are languages that have unique scripts and are more distantly related to the Indo-European languages making up the majority of the training dataset such as Hebrew ({\footnotesize\textsf{HE}}), Telugu ({\footnotesize\textsf{TE}}), Chinese ({\footnotesize\textsf{ZH}}), and Korean ({\footnotesize\textsf{KO}}). These differences could be due to a lack of transfer due to linguistic distance, our byte level BPE tokenizer being a poor match for these languages, or variations in data quality.

\subsection{Translation}\label{subsec:cross-lingual}

\begin{figure}[t]
\begin{center}
\centerline{\includegraphics[width=1.0\columnwidth]{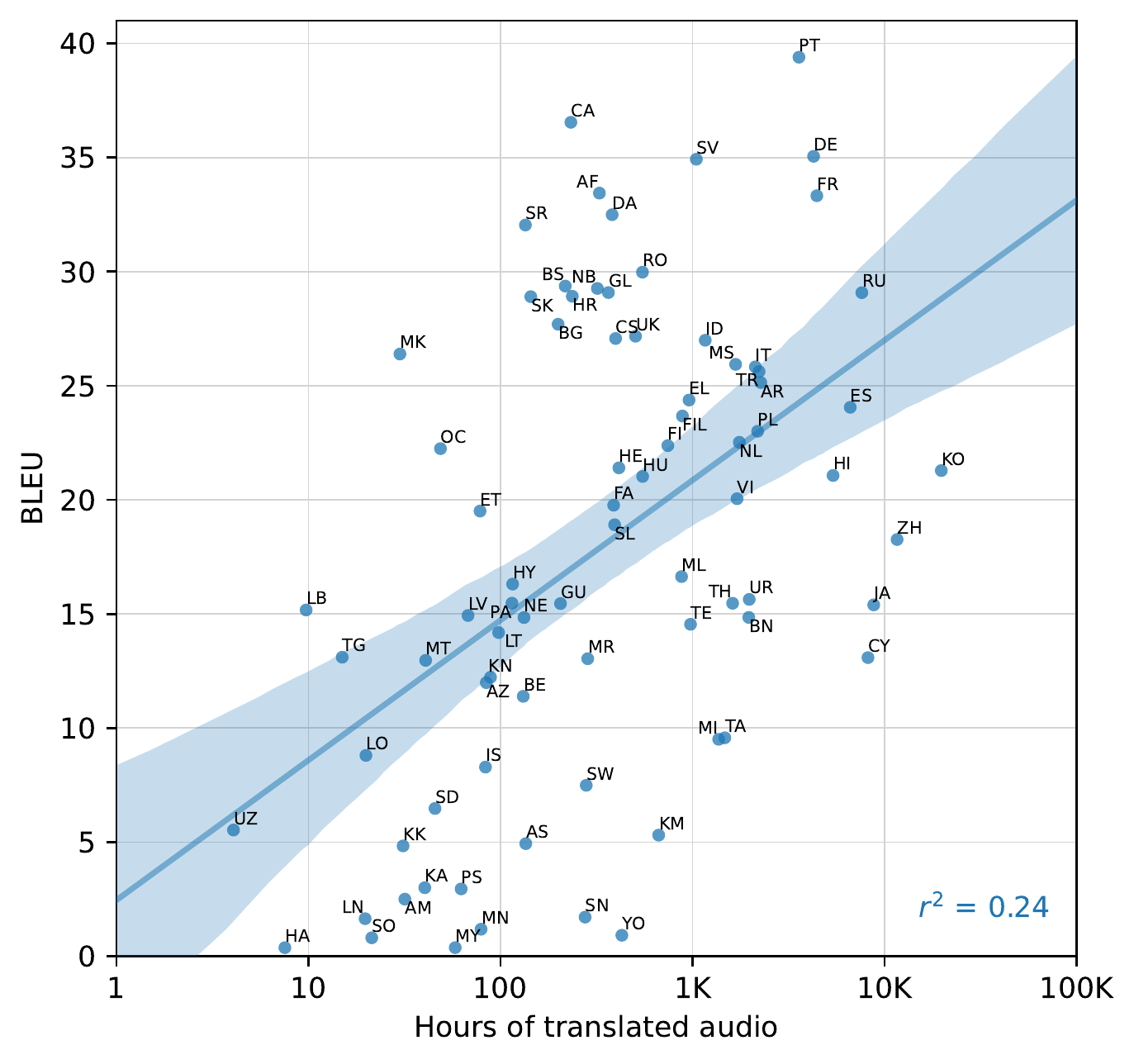}}
\caption{\textbf{Correlation of pre-training supervision amount with downstream translation performance.} The amount of pre-training translation data for a given language is only moderately predictive of Whisper's zero-shot performance on that language in Fleurs.}
\label{translation_transfer}
\end{center}
\vspace{-1em}
\end{figure}

\begin{table}[t]
\vskip 0.15in
\begin{center}
\begin{tabular}{l|ccc|c}
\toprule
 X → English & High & Mid & Low & All \\
\midrule
XMEF-X & 34.2 & 20.2 & 5.9 & 14.7 \\
XLS-R (2B) & 36.1 & 27.7 & 15.1 & 22.1 \\
mSLAM-CTC (2B) & 37.8 & 29.6 & 18.5 & 24.8 \\
Maestro & \textbf{38.2} & 31.3 & 18.4 & 25.2 \\
\midrule
Zero-Shot Whisper & 36.2 & \textbf{32.6} & \textbf{25.2} & \textbf{29.1} \\
\bottomrule
\end{tabular}
\caption{\textbf{\texttt{X$\rightarrow$en} Speech translation performance.} Zero-shot Whisper outperforms existing models on CoVoST2 in the overall, medium, and low resource settings but still moderately underperforms on high-resource languages compared to prior directly supervised work.}
\label{translate_table}
\end{center}
\vspace{-1em}
\end{table}

\begin{table}[t]
\vskip 0.15in
\begin{center}
\begin{tabular}{l|cc}
\toprule
 Language ID & Fleurs \\
\midrule
w2v-bert-51 (0.6B) & 71.4 \\
mSLAM-CTC (2B) & \textbf{77.7} \\
\midrule
Zero-shot Whisper & 64.5 \\
\bottomrule
\end{tabular}
\caption{\textbf{Language identification performance.} Zero-shot Whisper's accuracy at language identification is not competitive with prior supervised results on Fleurs. This is partially due to Whisper being heavily penalized for having no training data for 20 of Fleurs languages.}
\label{lang_id_table}
\end{center}
\vspace{-1em}
\end{table}

We study the translation capabilities of Whisper models by measuring their performance on the \texttt{X$\rightarrow$en} subset of CoVoST2 \cite{wang2020covost}. We compare with Maestro, mSLAM, and XLS-R, the highest-performing prior work. We achieve a new state of the art of 29.1 BLEU zero-shot without using any of the CoVoST2 training data. We attribute this to the 68,000 hours of \texttt{X$\rightarrow$en} translation data for these languages in our pre-training dataset which, although noisy, is vastly larger than the 861 hours of training data for \texttt{X$\rightarrow$en} translation in CoVoST2. Since Whisper evaluation is zero-shot, it does particularly well on the lowest resource grouping of CoVoST2, improving over mSLAM by 6.7 BLEU. Conversely, the best Whisper model does not actually improve over Maestro and mSLAM on average for the highest resource languages.

For an additional analysis on an even wider set of languages, we also re-purpose Fleurs, which is a speech recognition dataset, as a translation dataset. Since the same sentences are transcribed for every language we use the English transcripts as reference translations. In Figure \ref{translation_transfer} we visualize the correlation between the amount of translation training data per language and the resulting zero-shot BLEU score on Fleurs. While there is a clear trend of improvement with increasing training data, the squared correlation coefficient is much lower than the 0.83 observed for speech recognition and only 0.24. We suspect this is partly caused by the noisier training data due to errors in audio language identification. As an example, Welsh ({\footnotesize\textsf{CY}}) is an outlier with much worse than expected performance at only 13 BLEU despite supposedly having 9,000 hours of translation data. This large amount of Welsh translation data is surprising, ranking 4th overall for translation data and ahead of some of the most spoken languages in the world like French, Spanish, and Russian. Inspection shows the majority of supposedly Welsh translation data is actually English audio with English captions where the English audio was mis-classified as Welsh by the language identification system, resulting in it being included as translation training data rather transcription data according to our dataset creation rules.

\subsection{Language Identification}\label{subsec:language-id}

\begin{figure}[t]
\begin{center}
\centerline{\includegraphics[width=1.0\columnwidth]{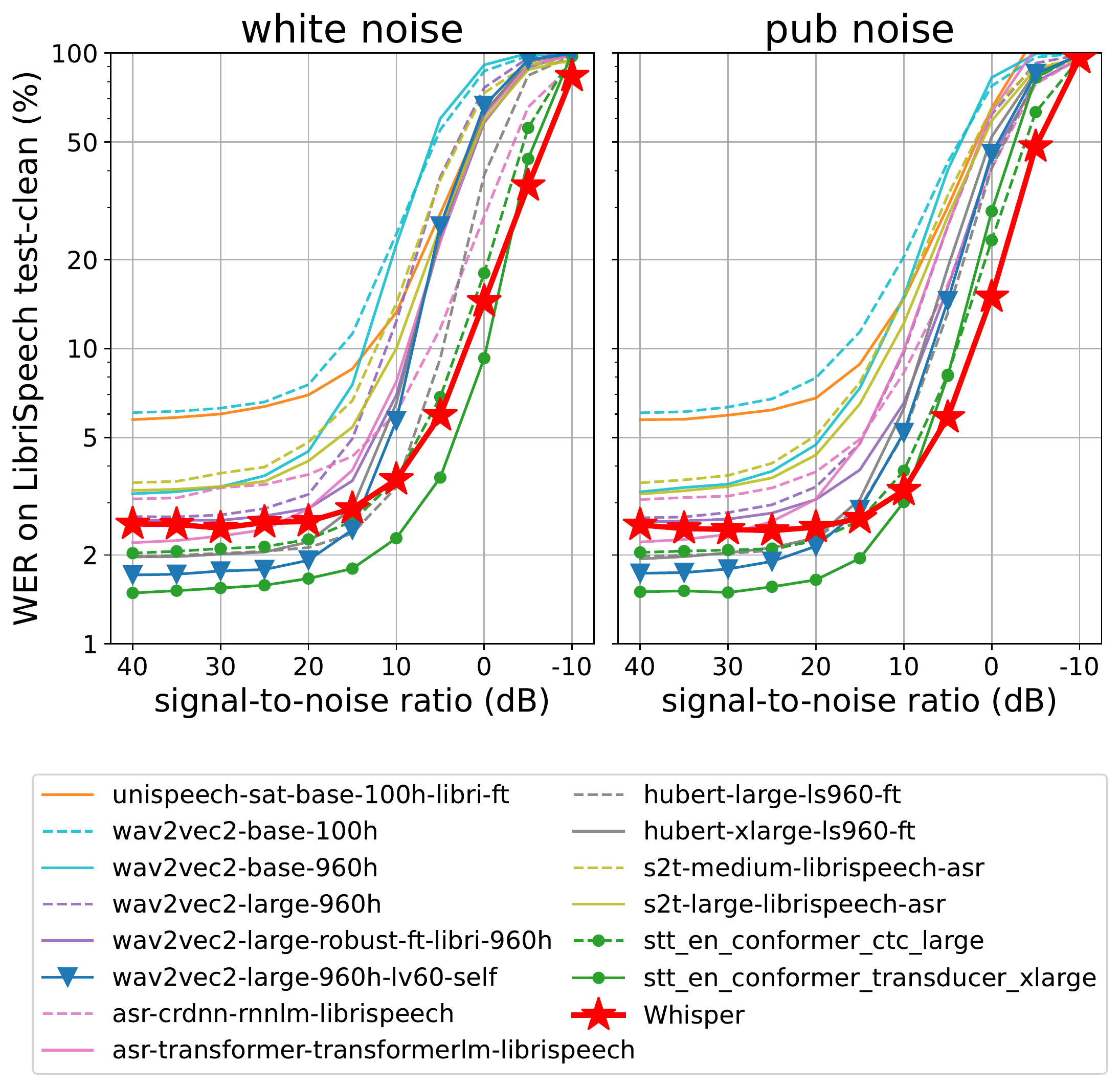}}
\definecolor{C0}{RGB}{31, 119, 180}
\definecolor{C2}{RGB}{44, 160, 44}
\definecolor{C4}{RGB}{214, 39, 40}
\caption{\textbf{WER on LibriSpeech test-clean as a function of SNR under additive white noise (left) and pub noise (right).} The accuracy of LibriSpeech-trained models degrade faster than the best Whisper model ({\color{red}$\bigstar$}). NVIDIA STT models ({\color{C2}$\bullet$}) perform best under low noise but are outperformed by Whisper under high noise (SNR $<$ 10 dB). The second-best model under low noise ({\color{C0}$\blacktriangledown$}) is fine-tuned on LibriSpeech only and degrades even more quickly.}
\label{fig:noise-robustness}
\end{center}
\vspace{-1em}
\end{figure}

\begin{figure*}[b]
\begin{center}
\centerline{\includegraphics[width=1.0\textwidth]{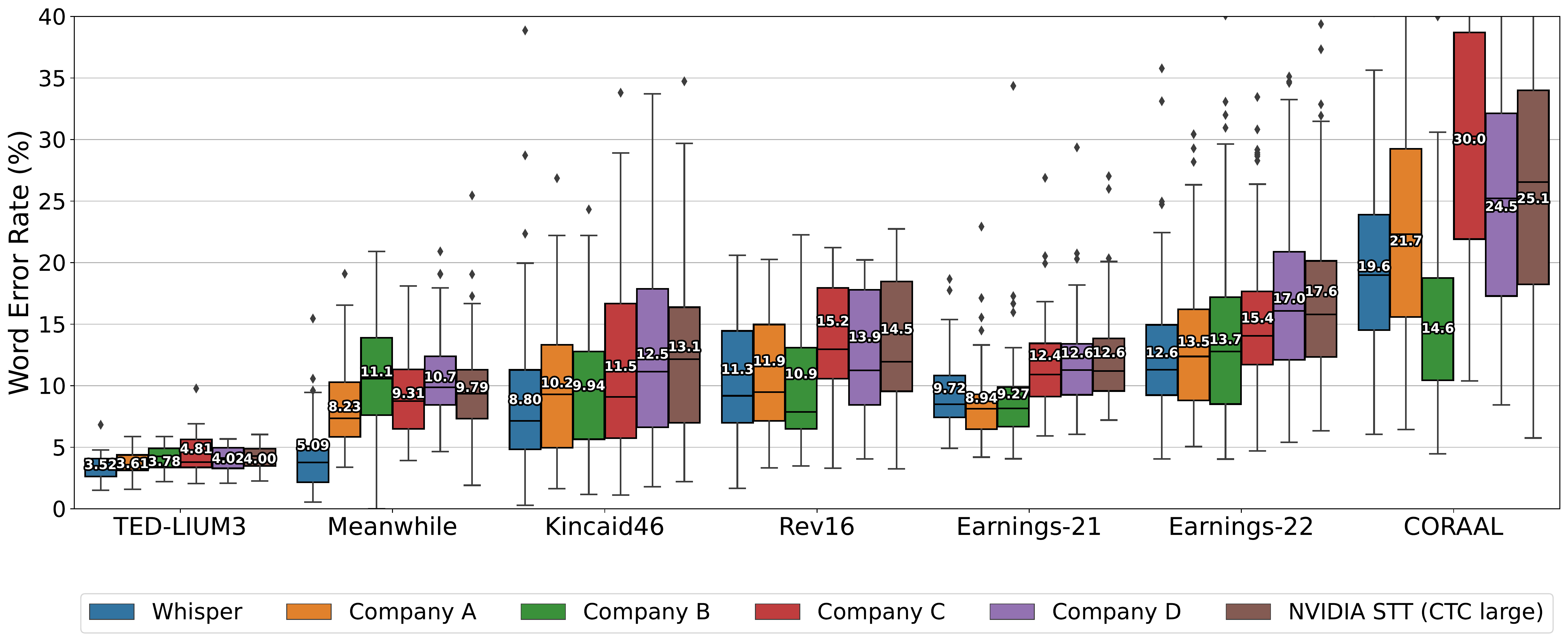}}
\caption{\textbf{Whisper is competitive with state-of-the-art commercial and open-source ASR systems in long-form transcription.} The distribution of word error rates from six ASR systems on seven long-form datasets are compared, where the input lengths range from a few minutes to a few hours. The boxes show the quartiles of per-example WERs, and the per-dataset aggregate WERs are annotated on each box. Our model outperforms the best open source model (NVIDIA STT) on all datasets, and in most cases, commercial ASR systems as well.}
\label{fig:long-form}
\end{center}
\vspace{-1em}
\end{figure*}

To evaluate language identification, we use the Fleurs dataset \cite{conneau2022fleurs}. The zero-shot performance of Whisper is not competitive with prior supervised work here and underperforms the supervised SOTA by 13.6\%. However, Whisper is heavily disadvantaged for language identification on Fleurs, since the Whisper dataset contains no training data for 20 of the 102 languages in Fleurs, upper-bounding accuracy at 80.4\%. On the 82 overlapping languages the best Whisper model achieves 80.3\% accuracy.

\subsection{Robustness to Additive Noise}\label{subsec:noise-robustness}

We tested the noise robustness of Whisper models and 14 LibriSpeech-trained models by measuring the WER when either white noise or pub noise from the Audio Degradation Toolbox \cite{mauch2013adt} was added to the audio. The pub noise represents a more natural noisy environment with ambient noise and indistinct chatter typical in a crowded restaurant or a pub. Among the 14 models, twelve are pre-trained and/or fine-tuned on LibriSpeech, and the other two are NVIDIA STT models trained on a mixture dataset similar to prior work like SpeechStew that includes LibriSpeech.
The level of additive noise corresponding to a given signal-to-noise ratio (SNR) is calculated based on the signal power of individual examples.
Figure \ref{fig:noise-robustness} shows how the ASR performance degrades as the additive noise becomes more intensive. There are many models that outperform our zero-shot performance under low noise (40 dB SNR), which is unsurprising given those models are trained primarily on LibriSpeech, but all models quickly degrade as the noise becomes more intensive, performing worse than the Whisper model under additive pub noise of SNR below 10 dB. This showcases Whisper's robustness to noise, especially under more natural distribution shifts like the pub noise.

\subsection{Long-form Transcription}\label{subsec:long-form}

Whisper models are trained on 30-second audio chunks and cannot consume longer audio inputs at once. This is not a problem with most academic datasets comprised of short utterances but presents challenges in real-world applications which often require transcribing minutes- or hours-long audio. We developed a strategy to perform buffered transcription of long audio by consecutively transcribing 30-second segments of audio and shifting the window according to the timestamps predicted by the model. We observed that it is crucial to have beam search and temperature scheduling based on the repetitiveness and the log probability of the model predictions in order to reliably transcribe long audio. The full procedure is described in Section \ref{subsec:long-form-strategy}.

We evaluate the long-form transcription performance on seven datasets consisting of speech recordings of various lengths and recording conditions, to cover as diverse a data distribution as possible. These include a long-form adaptation of TED-LIUM3 \cite{Hernandez2018TEDLIUM3T} concatenated so that each example is a full-length TED talk, a collection of jargon-laden segments taken from The Late Show with Stephen Colbert (Meanwhile), sets of videos/podcasts that has been used as ASR benchmarks in online blogs (Rev16 and Kincaid46), recordings of earnings calls \cite{del2021earnings}, and the full-length interviews from the Corpus of Regional African American Language (CORAAL) \cite{gunter2021contextualizing}. Full details about the long-form datasets can be found in Appendix \ref{sec:datasets}.

We compare the performance with open-source models as well as 4 commercial ASR services. The results are summarized in Figure \ref{fig:long-form}, showing the distribution of word error rates from Whisper and the 4 commercial ASR services, as well as the NVIDIA STT Conformer-CTC Large model from the NeMo toolkit \cite{kuchaiev2019nemo} which performed the best among the open-source models. All commercial ASR services are queried using their default English transcription settings as of September 1st, 2022, and for the NVIDIA STT model we used their buffered inference implementation in the \texttt{FrameBatchASR} class to enable long-form transcription. The results show that Whisper performs better than the compared models on most datasets, especially on the Meanwhile dataset which is heavy with uncommon words. Additionally, we note the possibility that some of the commercial ASR systems have been trained on some of these publicly available datasets, and therefore these results may not be accurately reflecting the relative robustness of the systems.

\subsection{Comparison with Human Performance}\label{subsec:human-comparison}

\begin{figure}[t]
\begin{center}
\centerline{\includegraphics[width=\columnwidth]{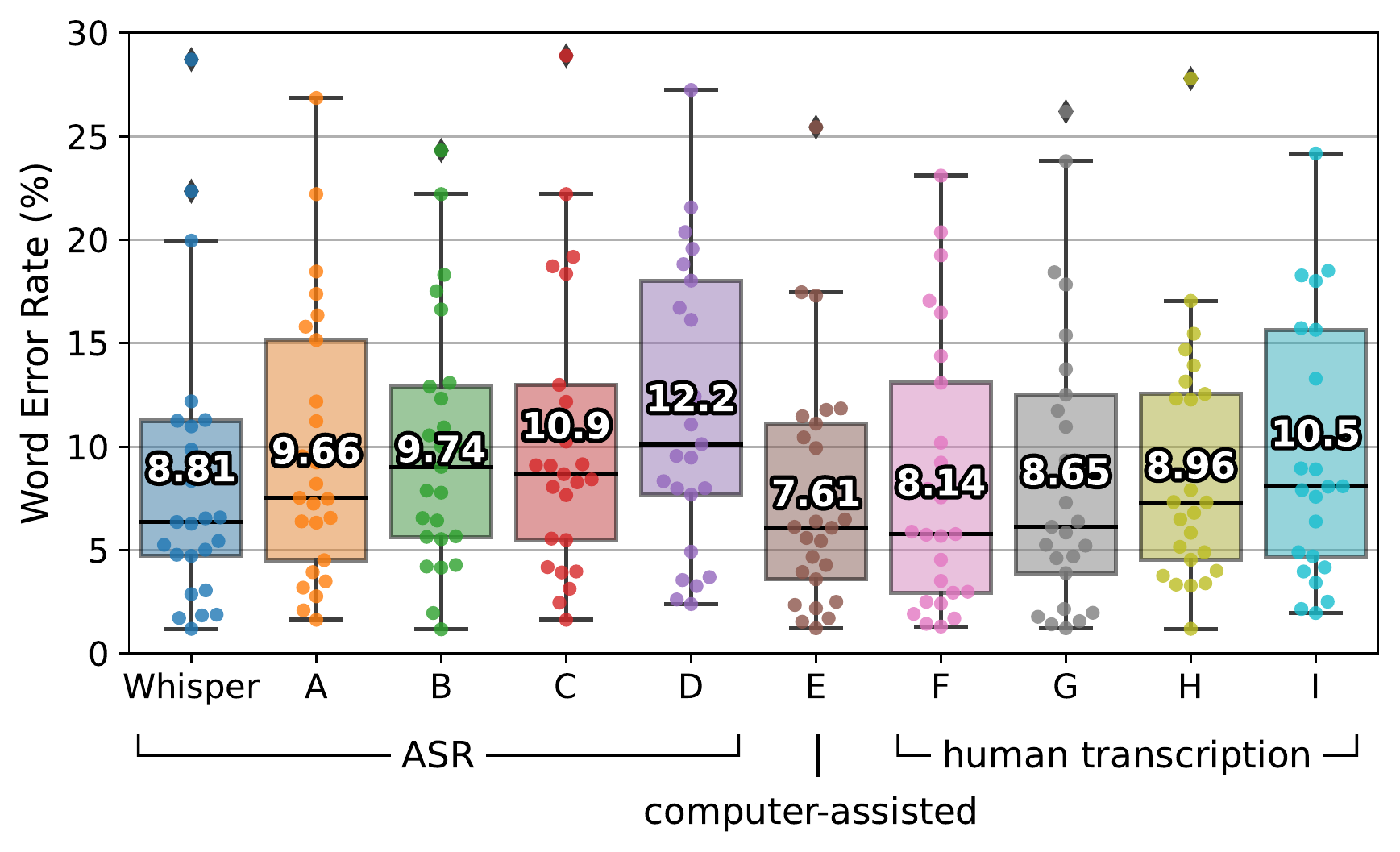}}
\caption{\textbf{Whisper's performance is close to that of professional human transcribers.} This plot shows the WER distributions of 25 recordings from the Kincaid46 dataset transcribed by Whisper, the same 4 commercial ASR systems from Figure \ref{fig:long-form} (A-D), one computer-assisted human transcription service (E) and 4 human transcription services (F-I). The box plot is superimposed with dots indicating the WERs on individual recordings, and the aggregate WER over the 25 recordings are annotated on each box.}
\label{fig:human-evals}
\end{center}
\vspace{-1em}
\end{figure}

Because of ambiguous or indistinct speech as well as labeling errors, there are different levels of irreducible error in each dataset, and with WER metrics from ASR systems alone it is difficult to make sense of how much room for improvement exists in each dataset. To quantify how close Whisper's performance is to the human performance, we selected 25 recordings from the Kincaid46 dataset and used 5 services to obtain transcripts produced by professional transcribers, among which one provides computer-assisted transcription and the other four are entirely human-transcribed. The audio selection covers various recording conditions such as scripted and unscripted broadcast, telephone and VoIP calls, and meetings. Figure \ref{fig:human-evals} shows the distribution of per-example WERs and aggregate WER across the 25 recordings, where the computer-assisted service has the lowest aggregate WER that is 1.15\% point better than Whisper's, and the pure-human performance is only a fraction of a percentage point better than Whisper's. These results indicate that Whisper's English ASR performance is not perfect but very close to human-level accuracy.

\section{Analysis and Ablations}\label{sec:ablation}

\subsection{Model Scaling}

\begin{figure*}[t]
\begin{center}
\centerline{\includegraphics[width=\textwidth]{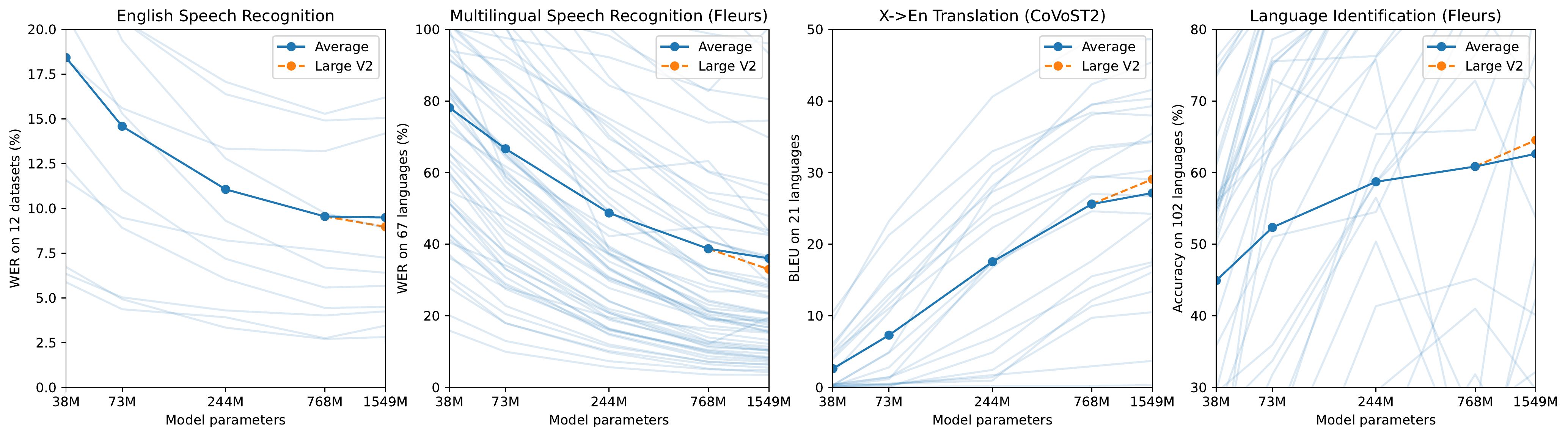}}
\caption{\textbf{Zero-shot Whisper performance scales reliably across tasks and languages with increasing model size.} Lightly shaded lines represent individual datasets or languages, showing that performance is more varied than the smooth trends in aggregate performance. Large V2 distinguished with a dashed orange line since it includes several changes that are not present for the smaller models in this analysis.}
\label{scaling}
\end{center}
\vspace{-1em}
\end{figure*}

A large amount of the promise in weakly supervised training approaches is their potential to use datasets much larger than those in traditional supervised learning. However, this comes with the cost of using data that is possibly much noisier and lower quality than gold-standard supervision. A concern with this approach is that although it may look promising to begin with, the performance of models trained on this kind of data may saturate at the inherent quality level of the dataset, which could be far below human level. A related concern is that as capacity and compute spent training on the dataset increases, models may learn to exploit the idiosyncrasies of the dataset, and their ability to generalize robustly to out-of-distribution data could even degrade. 

To check whether this is the case, we study the zero-shot generalization of Whisper models as a function of the model size. Our analysis is summarized in Figure \ref{scaling}. With the exception of English speech recognition, performance continues to increase with model size across multilingual speech recognition, speech translation, and language identification. The diminishing returns for English speech recognition could be due to saturation effects from approaching human-level performance as analysis in Section \ref{subsec:human-comparison} suggests.

\subsection{Dataset Scaling}

At 680,000 hours of labeled audio, the Whisper dataset is one of the largest ever created in supervised speech recognition. Exactly how important is the raw dataset size to Whisper's performance? To study this, we trained a series of medium-sized models on subsampled versions of the dataset which are 0.5\%, 1\%, 2\%, 4\%, and 8\% of the full dataset size and compared their performance with the same medium-sized model trained on the whole dataset. Early stopping based on the validation loss was used to select model checkpoints for each dataset size. Evaluation was performed on an exponential moving average estimate of the parameters \cite{polyak1992acceleration} using a smoothing rate of 0.9999 to help reduce the effect of the learning rate not fully decaying to zero for the models trained on the subsampled datasets due to early stopping. Performance on English and multilingual speech recognition and \texttt{X$\rightarrow$en} translation is reported in Table \ref{dataset_scaling_table}.

All increases in the dataset size result in improved performance on all tasks, although we see significant variability in improvement rates across tasks and sizes. Performance improves rapidly on English speech recognition from 3,000 to 13,000 hours and then slows down noticeably between 13,000 and 54,000 hours. Using the full dataset, which corresponds to another 12.5$\times$ increase in size results in only a further 1 point drop in WER. This mirrors the diminishing returns observed with model size scaling for English speech recognition and could similarly be explained by saturation effects when approaching human-level performance.

\begin{table}[t]
\vskip 0.15in
\begin{center}
\begin{tabular}{l|ccc}
\toprule
 Dataset & English & Multilingual & X$\rightarrow$En \\
 size & WER ($\downarrow$) & WER ($\downarrow$) & BLEU ($\uparrow$) \\
\midrule
3405 & 30.5 & 92.4 & 0.2 \\
6811 & 19.6 & 72.7 & 1.7 \\
13621 & 14.4 & 56.6 & 7.9 \\
27243 & 12.3 & 45.0 & 13.9 \\
54486 & 10.9 & 36.4 & 19.2 \\
681070 & \textbf{9.9} & \textbf{29.2} & \textbf{24.8} \\
\bottomrule
\end{tabular}
\caption{\textbf{Performance improves with increasing dataset size. } English speech recognition performance refers to an average over 12 datasets while the Multilingual speech recognition reports performance on the overlapping subset of languages in Fleurs and \texttt{X$\rightarrow$en} translation reports average BLEU on CoVoST2. Dataset size reported in hours.}
\label{dataset_scaling_table}
\end{center}
\vspace{-1em}
\end{table}

Improvements in WER follow a power-law trend for multilingual speech recognition till 54,000 hours and then deviate from this trend, improving only a further 7 points when increasing to the full dataset size. For \texttt{X$\rightarrow$en} translation, performance is practically zero when training on 7,000 hours of audio or less, and then follows a roughly log-linear improvement trend till 54,000 hours before also showing diminishing returns when further scaling to the full dataset size.

The general trend across tasks of diminishing returns when moving from 54,000 hours to our full dataset size of 680,000 hours could suggest that the current best Whisper models are under-trained relative to dataset size and performance could be further improved by a combination of longer training and larger models. It could also suggest that we are nearing the end of performance improvements from dataset size scaling for speech recognition. Further analysis is needed to characterize ``scaling laws'' for speech recognition in order to decided between these explanations.

\subsection{Multitask and Multilingual Transfer}

A potential concern with jointly training a single model on many tasks and languages is the possibility of negative transfer where interference between the learning of several tasks results in performance worse than would be achieved by training on only a single task or language. To investigate whether this is occurring, we compared the performance of models trained on just English speech recognition with our standard multitask and multilingual training setup and measured their average performance across our suite of zero-shot English speech recognition benchmarks. We adjust for the amount of FLOPs spent training on the task of English speech recognition as only 65\% of compute is spent on this task in a joint training setup; analysis would otherwise be confounded by under-training on the task when compared to a same-sized English-only model.

\begin{figure}[t]
\begin{center}
\centerline{\includegraphics[width=1.0\columnwidth]{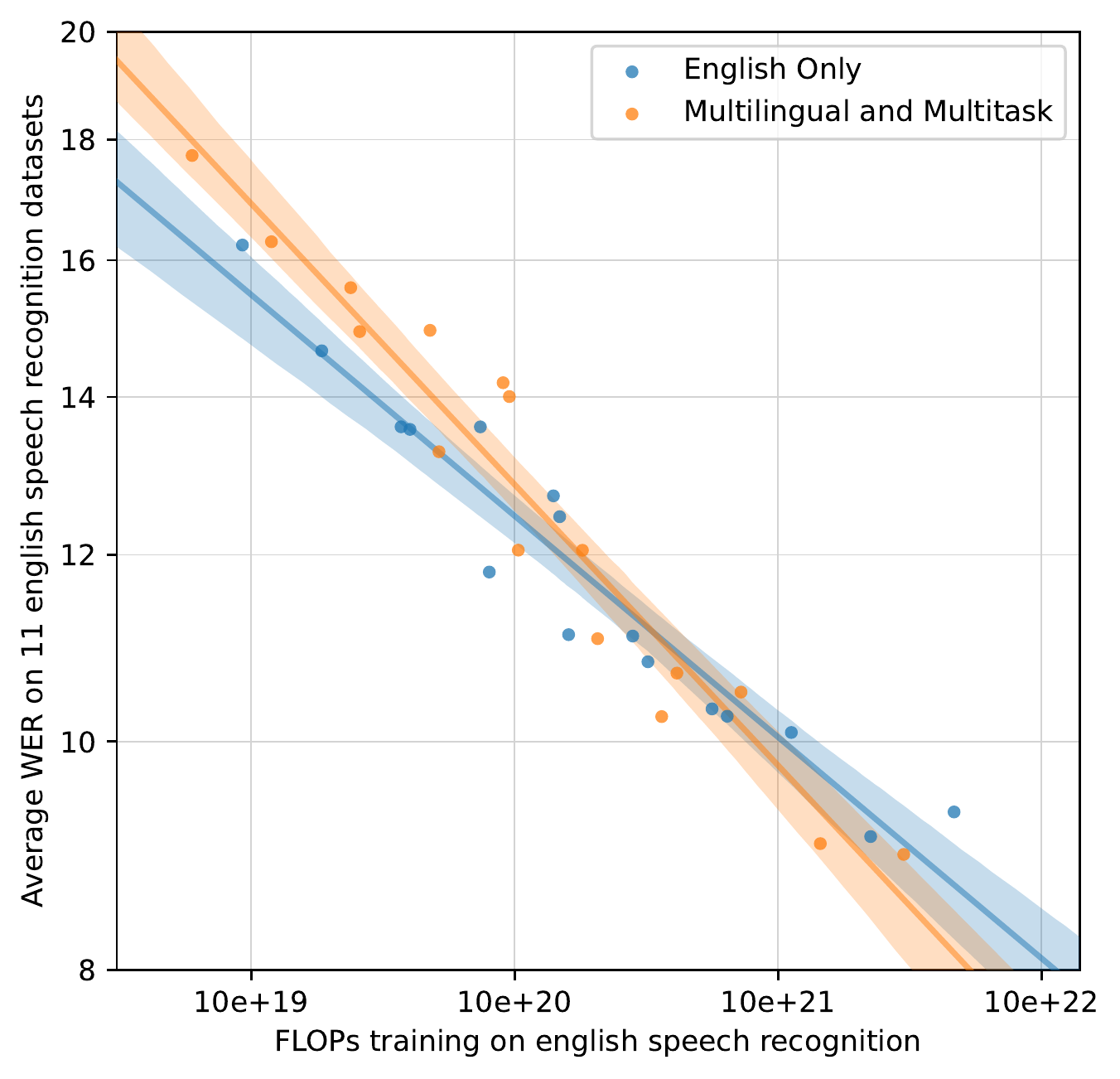}}
\caption{\textbf{Multitask and multilingual transfer improves with scale.} For small models, performance on English speech recognition degrades when trained jointly in a multitask and multilingual setup. However, multilingual and multitask models benefit more from scale and eventually outperform models trained on English data only. 95\% bootstrap estimate confidence intervals are shown.}
\label{multilingual_multitask_transfer}
\end{center}
\vspace{-1em}
\end{figure}

Our results visualized in Figure \ref{multilingual_multitask_transfer} show that for small models trained with moderate amounts of compute, there is indeed negative transfer between tasks and languages: joint models underperform English-only models trained for the same amount of compute. However, multitask and multilingual models scale better and for our largest experiments outperform their English-only counterparts demonstrating positive transfer from other tasks. For our largest experiments, joint models also slightly outperform English-only models even when not adjusting for compute spent per task.

\subsection{Text Normalization}\label{subsec:text-normalization-analysis}

\begin{figure}[t]
\begin{center}
\centerline{\includegraphics[width=1.0\columnwidth]{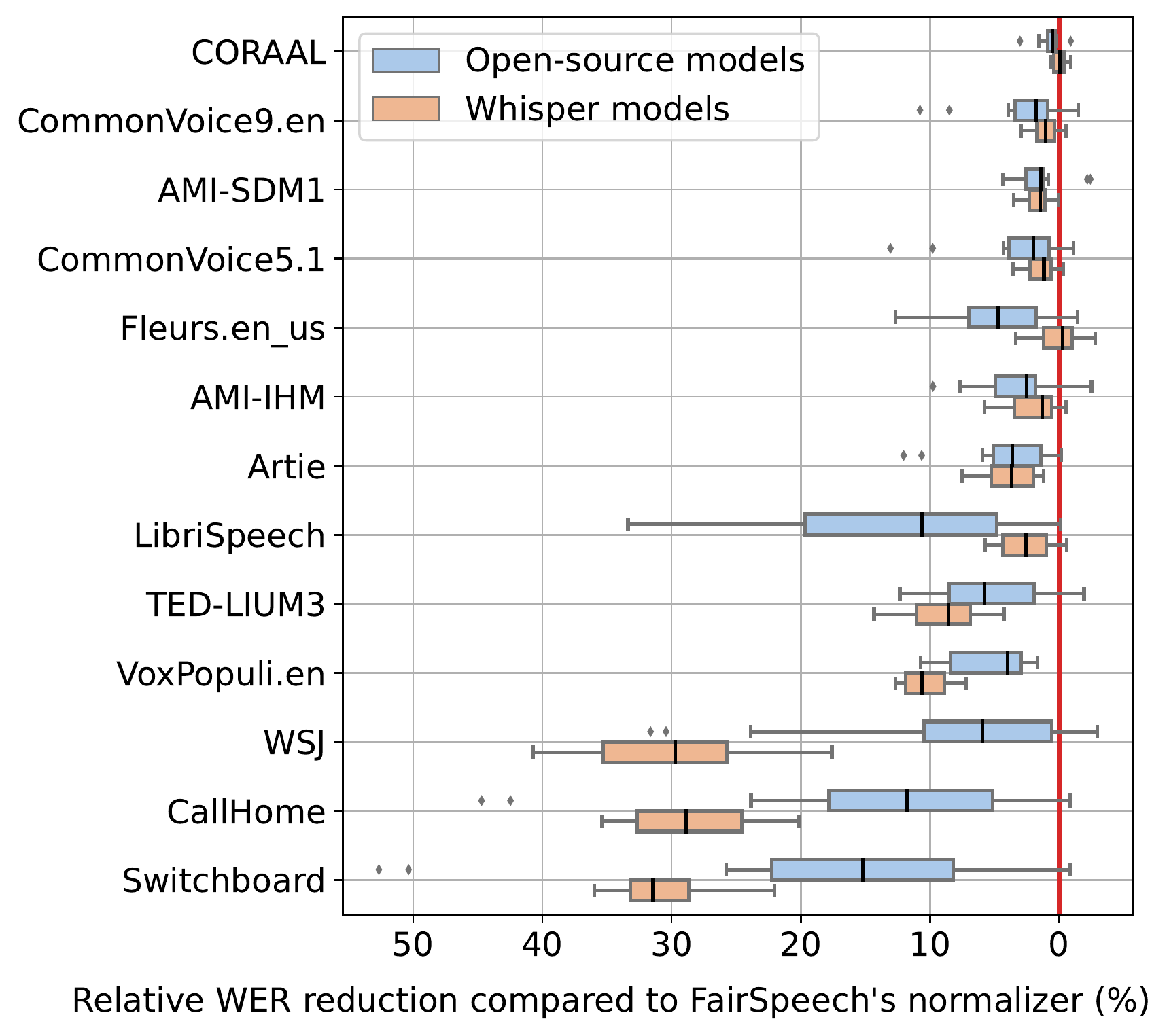}}
\caption{\textbf{On most datasets, our text normalizer has similar effect on reducing WERs between Whisper models and other open-source models, compared to FairSpeech's normalizer.} For each dataset, the boxplot shows the distribution of relative WER reduction across different models in our eval suite, showing that using our text normalizer generally results in lower WERs than FairSpeech's. On a few datasets our normalizer reduces WER significantly and more so for Whisper models, such as CallHome and Switchboard which have many contractions in the ground truth and WSJ which contains many numerical expressions.}
\label{fig:text_normalizer}
\end{center}
\vspace{-2em}
\end{figure}

Since we developed our text normalization jointly with Whisper to discount innocuous word errors, there is a risk that our normalizer is overfitted to fixing Whisper's peculiarities rather than addressing general variation in transcription. To check this, we compared the performance of Whisper using our normalizer versus an independently developed one from the FairSpeech project \cite{koenecke2020racial}. In Figure \ref{fig:text_normalizer}, we visualize the differences. On most datasets the two normalizers perform similarly, without significant differences in WER reduction between Whisper and compared open-source models, while on some datasets, namely WSJ, CallHome, and Switchboard, our normalizer reduces the WER of Whisper models' significantly more. The differences in reduction can be traced down to different formats used by the ground truth and how the two normalizers are penalizing them. For example, in CallHome and Switchboard, our standardizer did not penalize differences in common English contractions such as ``you're" versus ``you are", and in WSJ, our normalizer standardized the written and spoken forms of numerical and monetary expressions, such as ``sixty-eight million dollars" versus ``\$68 million".

\begin{table}[t]
\scriptsize
\setlength{\tabcolsep}{2pt}
\renewcommand{\arraystretch}{1.2}
\vskip 0.15in
\begin{center}

\begin{tabular}{lr|lcccccccr|lc}
\toprule
&&&

\rotatebox[origin=lc]{90}{TED-LIUM3~} &
\rotatebox[origin=lc]{90}{Meanwhile} &
\rotatebox[origin=lc]{90}{Kincaid46} &
\rotatebox[origin=lc]{90}{Rev16} &
\rotatebox[origin=lc]{90}{Earnings-21} &
\rotatebox[origin=lc]{90}{Earnings-22} &
\rotatebox[origin=lc]{90}{CORAAL} &

&&
\rotatebox[origin=lc]{90}{Average}\\
\midrule

Greedy decoding only &&& 3.95 & 5.16 & 9.69 & 11.7 & 10.7 & 14.0 & 22.0 &&& 11.0 \\
+ Beam search &&& 4.16 & 5.71 & 9.42 & 11.5 & 10.2 & 13.4 & 20.0 &&& 10.6 \\
+ Temperature fallback &&& 4.16 & 5.71 & 9.42 & 11.5 & 10.2 & 13.4 & 20.0 &&& 10.6 \\
+ Voice activity detection &&& 3.56 & \textbf{4.61} & 9.45 & 11.4 & 10.1 & 13.2 & 19.4 &&& 10.2 \\
+ Previous text conditioning &&& \textbf{3.42} & 6.16 & 8.72 & \textbf{11.0} & \textbf{9.63} & 13.3 & \textbf{18.1} &&& 10.0 \\
+ Initial timestamp constraint &&& 3.51 & 5.26 & \textbf{8.41} & 11.5 & 9.73 & \textbf{12.6} & 19.1 &&& 10.0 \\

\bottomrule
\end{tabular}

\caption{Long-form transcription performance improves incrementally as additional decoding heuristics are employed. Details on each intervention are described in Section \ref{subsec:long-form-strategy}.}
\label{tab:long-form-ablation}
\end{center}
\vspace{-1em}
\end{table}
\subsection{Strategies for Reliable Long-form Transcription}\label{subsec:long-form-strategy}

Transcribing long-form audio using Whisper relies on accurate prediction of the timestamp tokens to determine the amount to shift the model's 30-second audio context window by, and inaccurate transcription in one window may negatively impact transcription in the subsequent windows.
We have developed a set of heuristics that help avoid failure cases of long-form transcription, which is applied in the results reported in sections \ref{subsec:long-form} and \ref{subsec:human-comparison}.
First, we use beam search with 5 beams using the log probability as the score function, to reduce repetition looping which happens more frequently in greedy decoding. We start with temperature 0, i.e. always selecting the tokens with the highest probability, and increase the temperature by 0.2 up to 1.0 when either the average log probability over the generated tokens is lower than $-1$ or the generated text has a gzip compression rate higher than 2.4. Providing the transcribed text from the preceding window as previous-text conditioning when the applied temperature is below 0.5 further improves the performance.
We found that the probability of the \texttt{<|nospeech|>} token alone is not sufficient to distinguish a segment with no speech, but combining the no-speech probability threshold of 0.6 and the average log-probability threshold of $-1$ makes the voice activity detection of Whisper more reliable.
Finally, to avoid a failure mode where the model ignores the first few words in the input, we constrained the initial timestamp token to be between 0.0 and 1.0 second.
Table \ref{tab:long-form-ablation} shows that adding each of the interventions above incrementally reduces the WER overall, but not evenly across the dataset.
These heuristics serve as a workaround for the noisy predictions of the model, and more research would be needed to further improve the reliability of long-form decoding.

\section{Related Work}\label{sec:related}

\paragraph{Scaling Speech Recognition} A consistent theme across speech recognition research has been documenting the benefits of scaling compute, models, and datasets. Early work applying deep learning to speech recognition found improved performance with model depth and size and leveraged GPU acceleration to make training these larger models tractable \cite{mohamed2009deep}. Further research demonstrated that the benefit of deep learning approaches to speech recognition increased with dataset size, improving from being only competitive with prior GMM-HMM systems when using just 3 hours of TIMIT training data for phone recognition to achieving a 30\% word error rate reduction when trained on the 2,000 hour Switchboard dataset \cite{seide2011feature}. \citet{liao2013large} is an early example of leveraging weakly supervised learning to increase the size of a deep learning based speech recognition dataset by over 1,000 hours. These trends continued with Deep Speech 2 \cite{amodei2015deepspeech2} being a notable system developing high-throughput distributed training across 16 GPUs and scaling to 12,000 hours of training data while demonstrating continuing improvements at that scale. By leveraging semi-supervised pre-training, \citet{narayanan2018toward} were able to grow dataset size much further and study training on 162,000 hours of labeled audio. More recent work has explored billion-parameter models \cite{zhang2020pushing} and using up to 1,000,000 hours of training data \cite{zhang2021bigssl}.

\paragraph{Multitask Learning} Multitask learning \cite{caruana1997multitask} has been studied for a long time. In speech recognition, multi-lingual models have been explored for well over a decade \cite{schultz2006multilingual}. An inspirational and foundational work in NLP exploring multi-task learning with a single model is \citet{collobert2011natural}. Multitask learning in the sequence-to-sequence framework \cite{sutskever2014sequence} using multiple encoders and decoders was investigated in \citet{luong2015multi}. The use of language codes with a shared encoder/decoder architecture was first demonstrated for machine translation by \citet{johnson2017google}, removing the need for separate encoders and decoders. This approach was simplified further into the ``text-to-text'' framework of \citet{mccann2018natural} and popularized by its success with large transformer language models in the work of \citet{radford2019gpt2} and \citet{raffel2020exploring}. \citet{Toshniwal2018MultilingualSR} demonstrated jointly training a modern deep learning speech recognition system on several languages with a single model, and \citet{Pratap2020MassivelyMA} scaled this line of work significantly to 50 languages with a billion-parameter model. MUTE \cite{wang2020multitask} and mSLAM \cite{bapna2022mslam} studied joint training over both text and speech language tasks, demonstrating transfer between them.

\paragraph{Robustness} The question of how effectively models transfer and how robust they are to distribution shift and other types of perturbations has long been studied and is actively being researched across many fields of machine learning. \citet{torralba2011databias} highlighted the lack of generalization of machine learning models between datasets over a decade ago. Many other works have shown and continually reiterated how despite high performance on IID test sets, machine learning models can still make many mistakes when evaluated in even slightly different settings \cite{lake2017building,jia2017adversarial,alcorn2019strike,barbu2019objectnet,recht2019generalize}. More recently, \citet{taori2020robustness} studied the robustness of image classification models, and \citet{miller2020nlprobustness} investigated this for question-answering models. A key finding has been that multi-domain training increases robustness and generalization as discussed in the Introduction. This finding has been replicated across many fields in addition to speech recognition including NLP \cite{hendrycks2020pretrained} and computer vision \cite{radford2021clip}.

\section{Limitations and Future Work}\label{sec:future}


From our experimental results, analyses, and ablations, we have noted several limitations and areas for future work.

\paragraph{Improved decoding strategies.} As we have scaled Whisper, we have observed that larger models have made steady and reliable progress on reducing perception-related errors such as confusing similar-sounding words. Many remaining errors, particularly in long-form transcription seem more stubborn in nature and decidedly non-human/perceptual. They are a combination of failure modes of seq2seq models, language models, and text-audio alignment and include problems such as getting stuck in repeat loops, not transcribing the first or last few words of an audio segment, or complete hallucination where the model will output a transcript entirely unrelated to the actual audio. Although the decoding details discussed in Section \ref{subsec:long-form-strategy} help significantly, we suspect fine-tuning Whisper models on a high-quality supervised dataset and/or using reinforcement learning to more directly optimize for decoding performance could help further reduce these errors.

\paragraph{Increase Training Data For Lower-Resource Languages} As Figure \ref{fleurs_asr_transfer} shows, Whisper's speech recognition performance is still quite poor on many languages. The same analysis suggests a clear route for improvement since performance on a language is very well predicted by the amount of training data for the language. Since our pre-training dataset is currently very English-heavy due to biases of our data collection pipeline, which sourced primarily from English-centric parts of the internet, most languages have less than 1000 hours of training data. A targeted effort at increasing the amount of data for these rarer languages could result in a large improvement to average speech recognition performance even with only a small increase in our overall training dataset size.

\paragraph{Studying fine-tuning} In this work, we have focused on the robustness properties of speech processing systems and as a result only studied the zero-shot transfer performance of Whisper. While this is a crucial setting to study due to it being representative of general reliability, for many domains where high-quality supervised speech data does exist, it is likely that results can be improved further by fine-tuning. An additional benefit of studying fine-tuning is that it allows for direct comparisons with prior work since it is a much more common evaluation setting.

\paragraph{Studying the impact of Language Models on Robustness} As argued in the introduction, we suspect that Whisper's robustness is partially due to its strong decoder, which is an audio conditional language model. It's currently unclear to what degree the benefits of Whisper stem from training its encoder, decoder, or both. This could be studied by either ablating various design components of Whisper, such as training a decoder-less CTC model, or by studying how the performance of existing speech recognition encoders such as wav2vec 2.0 change when used together with a language model.


\paragraph{Adding Auxiliary Training Objectives} Whisper departs noticeably from most recent state-of-the-art speech recognition systems due to the lack of unsupervised pre-training or self-teaching methods. While we have not found them necessary to achieve good performance, it is possible that the results could be further improved by incorporating this.

\section{Conclusion}\label{sec:conclusion}

Whisper suggests that scaling weakly supervised pre-training has been underappreciated so far in speech recognition research. We achieve our results without the need for the self-supervision and self-training techniques that have been a mainstay of recent large-scale speech recognition work and demonstrate how simply training on a large and diverse supervised dataset and focusing on zero-shot transfer can significantly improve the robustness of a speech recognition system.

\subsubsection*{Acknowledgments}
We'd like to thank the millions of people who were involved in creating the data used by Whisper. We'd also like to thank Nick Ryder, Will Zhuk, and Andrew Carr for the conversation on the waterfall hike that inspired this project. We are also grateful to the Acceleration and Supercomputing teams at OpenAI for their critical work on software and hardware infrastructure this project used. We'd also like to thank Pamela Mishkin for advising the project from a policy perspective. Finally, we are grateful to the developers of the many software packages used throughout this project including, but not limited, to Numpy \citep{2020NumPy-Array}, SciPy \citep{2020SciPy-NMeth}, ftfy \citep{speer-2019-ftfy}, PyTorch \citep{NEURIPS2019_9015}, pandas \citep{reback2020pandas}, and scikit-learn \citep{scikit-learn}.

\bibliography{asr_paper}
\bibliographystyle{icml2022}

\newpage
\appendix
\onecolumn

\section{Evaluation Datasets.}\label{sec:datasets}

\subsection{Short-form English-only datasets}

\begin{itemize}
    
\item \textbf{LibriSpeech} \cite{panayotov2015librispeech}: We used the test-clean and test-other splits from the \href{https://www.openslr.org/12}{LibriSpeech ASR corpus}.

\item \textbf{TED-LIUM 3} \cite{Hernandez2018TEDLIUM3T}: We used the test split of \href{https://www.openslr.org/51/}{TED-LIUM Release 3}, using the segmented manual transcripts included in the release.

\item \textbf{Common Voice 5.1} \cite{ardila2019common}: We downloaded the English subset of Common Voice Corpus 5.1 from \href{https://commonvoice.mozilla.org/en/datasets}{the official website}.

\item \textbf{Artie bias corpus} \cite{meyer2020artie}: We used the \href{https://github.com/artie-inc/artie-bias-corpus}{Artie bias corpus}. This is a subset of the Common Voice dataset.

\item \textbf{CallHome} and \textbf{Switchboard}: We used the two corpora from \href{https://catalog.ldc.upenn.edu/LDC2002S09}{LDC2002S09} and \href{https://catalog.ldc.upenn.edu/LDC2002T43}{LDC2002T43}.

\item \textbf{WSJ}: We used \href{https://catalog.ldc.upenn.edu/LDC93S6B}{LDC93S6B} and \href{https://catalog.ldc.upenn.edu/LDC94S13B}{LDC94S13B} and followed the \href{https://github.com/kaldi-asr/kaldi/tree/master/egs/wsj/s5}{\texttt{s5} recipe} to preprocess the dataset.

\item \textbf{CORAAL}: We used the 231 interviews from CORAAL \cite{kendall2021coraal} and used the preprocessing script from \href{https://github.com/stanford-policylab/asr-disparities/blob/master/input/CORAAL_transcripts.csv}{the FairSpeech project}.

\item \textbf{CHiME-6}: For CHiME-6 \cite{watanabe2020chime}, we downloaded the \href{https://spandh.dcs.shef.ac.uk//chime_challenge/CHiME5/download.html}{CHiME-5 dataset} and followed the stage 0 of the \href{https://github.com/kaldi-asr/kaldi/tree/master/egs/chime6/s5_track1}{\texttt{s5\_track1} recipe} to create the CHiME-6 dataset which fixes synchronization. We then used the binaural recordings (\texttt{*\_P??.wav}) and the corresponding transcripts.

\item \textbf{AMI-IHM} and \textbf{AMI-SDM1}: We preprocessed the \href{https://groups.inf.ed.ac.uk/ami/corpus/overview.shtml}{AMI Corpus} by following the stage 0 ad 2 of the \href{https://github.com/kaldi-asr/kaldi/tree/master/egs/ami/s5b}{s5b recipe}.

\end{itemize}

\subsection{Long-form English-only datasets}

\begin{itemize}
\item \textbf{TED-LIUM 3} \cite{Hernandez2018TEDLIUM3T}: We used the 11 full-length TED talks from the test split of \href{https://www.openslr.org/51/}{TED-LIUM Release 3}, slicing the source audio files between the beginning of the first labeled segment and the end of the last labeled segment of each talk, and we used the concatenated text as the label.

\item \textbf{Meanwhile}: This dataset consists of 64 segments from The Late Show with Stephen Colbert. The YouTube video ID and the corresponding start and end timestamps are available as part of the code release. The labels are collected from the closed-caption data for each video and corrected with manual inspection.

\item \textbf{Rev16}: We use a subset of 16 files from the 30 podcast episodes in \href{https://www.rev.ai/blog/podcast-transcription-benchmark-part-1/}{Rev.AI's Podcast Transcription Benchmark}, after finding that there are multiple cases where a significant portion of the audio and the labels did not match, mostly on the parts introducing the sponsors. We selected 16 episodes that do not have this error, whose ``file number''s are:

\begin{center}
    \texttt{3 4 9 10 11 14 17 18 20 21 23 24 26 27 29 32}    
\end{center}

\item \textbf{Kincaid46}: This dataset consists of 46 audio files and the corresponding transcripts compiled in the blog article \href{https://medium.com/descript/which-automatic-transcription-service-is-the-most-accurate-2018-2e859b23ed19}{<Which automatic transcription service is the most accurate - 2018>} by Jason Kincaid. We used the 46 audio files and reference transcripts from the Airtable widget in the article. For the human transcription benchmark in the paper, we use a subset of 25 examples from this data, whose ``Ref ID''s are:

\begin{center}
    \texttt{2 4 5 8 9 10 12 13 14 16 19 21 23 25 26 28 29 30 33 35 36 37 42 43 45}
\end{center}

\item \textbf{Earnings-21} \cite{del2021earnings} and \textbf{Earnings-22}: We used the files available in the \href{https://github.com/revdotcom/speech-datasets}{speech-datasets repository}, as of their 202206 version.

\item \textbf{CORAAL}: We used the 231 full-length interviews and transcripts from \cite{kendall2021coraal}.

\end{itemize}

\subsection{Multilingual datasets}

\begin{itemize}
    \item \textbf{Multilingual LibriSpeech} \cite{pratap2020mls}: We used the test splits from each language in \href{https://www.openslr.org/94/}{the Multilingual LibriSpeech (MLS) corpus}.
    \item \textbf{Fleurs} \cite{conneau2022fleurs}: We collected audio files and transcripts using the implementation available as \href{https://huggingface.co/datasets/google/fleurs/blob/main/fleurs.py}{HuggingFace datasets}. To use as a translation dataset, we matched the numerical utterance IDs to find the corresponding transcript in English.
    \item \textbf{VoxPopuli} \cite{wang2021voxpopuli}: We used the \texttt{get\_asr\_data.py} script from \href{https://github.com/facebookresearch/voxpopuli}{the official repository} to collect the ASR data in 16 languages, including English.
    \item \textbf{Common Voice 9} \cite{ardila2019common}: We downloaded the Common Voice Corpus 9 from \href{https://commonvoice.mozilla.org/en/datasets}{the official website}.
    \item \textbf{CoVOST 2} \cite{wang2020covost}: We collected the X into English data collected using the official repository.
\end{itemize}

\section{Compared Models}\label{sec:baseline-models}

For comparison, we use the following models from HuggingFace, downloaded as of September 2022 using version 4.21.0 of the \texttt{transformers} library:

\begin{itemize}
    \item \texttt{facebook/wav2vec2-large-960h-lv60-self} \cite{xu2021complementary}
    \item \texttt{facebook/wav2vec2-large-robust-ft-libri-960h} \cite{hsu2021robust}
    \item \texttt{facebook/wav2vec2-base-100h} \cite{baevski2020wav2vec2}
    \item \texttt{facebook/wav2vec2-base-960h} \cite{baevski2020wav2vec2}
    \item \texttt{facebook/wav2vec2-large-960h} \cite{baevski2020wav2vec2}
    \item \texttt{facebook/hubert-large-ls960-ft} \cite{hsu2021hubert}
    \item \texttt{facebook/hubert-xlarge-ls960-ft} \cite{hsu2021hubert}
    \item \texttt{facebook/s2t-medium-librispeech-asr} \cite{wang2020fairseq}
    \item \texttt{facebook/s2t-large-librispeech-asr} \cite{wang2020fairseq}
    \item \texttt{microsoft/unispeech-sat-base-100h-libri-ft} \cite{chen2022unispeech}
    \item \texttt{nvidia/stt\_en\_conformer\_ctc\_large} \cite{kuchaiev2019nemo}
    \item \texttt{nvidia/stt\_en\_conformer\_transducer\_xlarge} \cite{kuchaiev2019nemo}
    \item \texttt{speechbrain/asr-crdnn-rnnlm-librispeech} \cite{speechbrain}
    \item \texttt{speechbrain/asr-transformer-transformerlm-librispeech} \cite{speechbrain}
\end{itemize}

We note that all of the models above are entirely or partly trained on LibriSpeech.

\vfill
\pagebreak

\section{Text Standardization}\label{sec:standardization}

Since Whisper may output any UTF-8 string rather than a restricted set of graphemes, the rules for text standardization need to be more intricate and comprehensive than those defined on e.g. ASCII characters. We perform the following steps to normalize English texts in different styles into a standardized form, which is a best-effort attempt to penalize only when a word error is caused by actually mistranscribing a word, and not by formatting or punctuation differences.

\begin{enumerate}
    \item Remove any phrases between matching brackets (\texttt{[}, \texttt{]}).
    \item Remove any phrases between matching parentheses (\texttt{(}, \texttt{)}).
    \item Remove any of the following words: \texttt{hmm}, \texttt{mm}, \texttt{mhm}, \texttt{mmm}, \texttt{uh}, \texttt{um}
    \item Remove whitespace characters that comes before an apostrophe \texttt{'}
    \item Convert standard or informal contracted forms of English into the original form.
    \item Remove commas (\texttt{,}) between digits
    \item Remove periods (\texttt{.}) not followed by numbers
    \item Remove symbols as well as diacritics from the text, where symbols are the characters with the Unicode category starting with \texttt{M}, \texttt{S}, or \texttt{P}, except period, percent, and currency symbols that may be detected in the next step.
    \item Detect any numeric expressions of numbers and currencies and replace with a form using Arabic numbers, e.g. ``Ten thousand dollars'' $\rightarrow$ ``\$10000''.
    \item Convert British spellings into American spellings.
    \item Remove remaining symbols that are not part of any numeric expressions.
    \item Replace any successive whitespace characters with a space.
\end{enumerate}

A different, language-specific set of transformations would be needed to equivalently normalize non-English text, but due to our lack of linguistic knowledge to build such normalizers for all languages, we resort to the following basic standardization for non-English text:

\begin{enumerate}
    \item Remove any phrases between matching brackets (\texttt{[}, \texttt{]}).
    \item Remove any phrases between matching parentheses (\texttt{(}, \texttt{)}).
    \item Replace any markers, symbols, and punctuation characters with a space, i.e. when the Unicode category of each character in the NFKC-normalized string starts with \texttt{M}, \texttt{S}, or \texttt{P}.
    \item make the text lowercase.
    \item replace any successive whitespace characters with a space.
\end{enumerate}

Additionally, we put a space between every letter for the languages that do not use spaces to separate words, namely Chinese, Japanese, Thai, Lao, and Burmese, effectively measuring the character error rate instead.

We note that the above is an imperfect solution, and it will sometimes produce unintended and unexpected outputs. We do not claim that the text format resulting from the above is more ``correct'' in any measure. Rather, the procedures above are designed to better distinguish between innocuous differences in wording and genuine mistranscriptions. Python code for the standardization procedures above is available as part of our code and model release to facilitate future iterations and improvements on text standardization.

\vfill
\pagebreak

\section{Raw Performance Tables}\label{sec:raw-performance}

\subsection{English Transcription}

\subsubsection{Greedy decoding}
\begin{table}[H]\centering
\scriptsize
\begin{tabular}{l|rrrrrrrrrrrrrr}
\toprule
\setlength{\tabcolsep}{2pt}\renewcommand{\arraystretch}{1.2}
Model
& \multicolumn{1}{c}{\rotatebox[origin=rc]{270}{LibriSpeech.test-clean}}
& \multicolumn{1}{c}{\rotatebox[origin=rc]{270}{LibriSpeech.test-other}}
& \multicolumn{1}{c}{\rotatebox[origin=rc]{270}{TED-LIUM3}}
& \multicolumn{1}{c}{\rotatebox[origin=rc]{270}{WSJ}}
& \multicolumn{1}{c}{\rotatebox[origin=rc]{270}{CallHome}}
& \multicolumn{1}{c}{\rotatebox[origin=rc]{270}{Switchboard}}
& \multicolumn{1}{c}{\rotatebox[origin=rc]{270}{CommonVoice5.1}}
& \multicolumn{1}{c}{\rotatebox[origin=rc]{270}{Artie}}
& \multicolumn{1}{c}{\rotatebox[origin=rc]{270}{CORAAL}}
& \multicolumn{1}{c}{\rotatebox[origin=rc]{270}{CHiME6}}
& \multicolumn{1}{c}{\rotatebox[origin=rc]{270}{AMI-IHM}}
& \multicolumn{1}{c}{\rotatebox[origin=rc]{270}{AMI-SDM1}}
& \multicolumn{1}{c}{\rotatebox[origin=rc]{270}{VoxPopuli.en}}
& \multicolumn{1}{c}{\rotatebox[origin=rc]{270}{Fleurs.en\_us}}
\\ \midrule
Whisper tiny.en 
& 5.6 & 14.6 & 6.0 & 5.0 & 24.1 & 17.8 & 26.3 & 20.0 & 23.9 & 41.3 & 23.7 & 50.3 & 11.7 & 11.6 
\\
Whisper tiny 
& 7.6 & 16.9 & 7.0 & 6.7 & 30.0 & 22.8 & 29.6 & 23.9 & 31.0 & 49.6 & 27.6 & 58.1 & 12.7 & 13.7 
\\
Whisper base.en 
& 4.2 & 10.2 & 4.9 & 4.6 & 20.9 & 15.2 & 19.0 & 13.4 & 22.6 & 36.4 & 20.5 & 46.7 & 10.0 & 7.6 
\\
Whisper base 
& 5.0 & 12.4 & 5.5 & 5.1 & 23.0 & 16.8 & 21.6 & 16.9 & 26.0 & 40.2 & 22.0 & 49.9 & 10.0 & 10.1 
\\
Whisper small.en 
& 3.1 & 7.4 & 4.0 & 3.3 & 18.2 & 15.7 & 13.1 & 9.7 & 20.2 & 27.6 & 17.5 & 38.0 & 8.1 & 6.0 
\\
Whisper small 
& 3.4 & 7.6 & 4.3 & 4.0 & 17.5 & 14.5 & 13.5 & 10.3 & 18.1 & 29.3 & 19.0 & 39.6 & 8.3 & 6.6 
\\
Whisper medium.en 
& 3.1 & 6.3 & 4.1 & 3.3 & 16.2 & 14.1 & 10.6 & 7.6 & 17.5 & 25.3 & 16.4 & 37.2 & 7.4 & 5.0 
\\
Whisper medium 
& 2.9 & 5.9 & 3.8 & 2.9 & 16.4 & 14.0 & 10.3 & 7.2 & 16.6 & 26.4 & 16.6 & 36.0 & 7.4 & 5.4 
\\
Whisper large 
& 2.7 & 5.6 & 4.0 & 3.1 & 15.8 & 13.1 & 9.5 & 6.7 & 19.4 & 25.6 & 16.4 & 36.9 & 7.3 & 4.6 
\\
Whisper large-v2
& 2.7 & 5.2 & 4.0 & 3.9 & 17.6 & 13.8 & 9.0 & 6.2 & 16.2 & 25.5 & 16.9 & 36.4 & 7.3 & 4.4
\\
\midrule
wav2vec2-base-100h 
& 6.0 & 13.4 & 17.8 & 13.9 & 46.9 & 40.2 & 47.4 & 40.8 & 47.0 & 79.9 & 48.1 & 81.2 & 28.9 & 23.1 
\\
wav2vec2-base-960h 
& 3.3 & 8.5 & 12.8 & 8.9 & 40.6 & 32.9 & 36.4 & 30.9 & 39.9 & 68.5 & 40.2 & 71.9 & 21.4 & 17.4 
\\
wav2vec2-large-960h-lv60-self 
& 1.8 & 3.8 & 7.4 & 4.4 & 29.1 & 22.2 & 19.9 & 15.8 & 29.2 & 56.3 & 30.8 & 57.0 & 13.0 & 10.2 
\\
wav2vec2-large-960h 
& 2.7 & 6.2 & 10.5 & 7.7 & 34.8 & 28.3 & 29.9 & 24.5 & 35.6 & 65.8 & 37.0 & 67.6 & 17.9 & 14.6 
\\
wav2vec2-large-robust-ft-libri-960h 
& 2.6 & 5.3 & 9.2 & 6.1 & 23.4 & 19.8 & 20.3 & 16.2 & 29.4 & 58.1 & 31.7 & 61.6 & 15.1 & 11.8 
\\
asr-crdnn-rnnlm-librispeech 
& 3.0 & 9.7 & 17.7 & 10.7 & 59.7 & 56.1 & 43.7 & 33.3 & 83.8 & 81.0 & 57.2 & 85.8 & 30.6 & 32.4 
\\
asr-transformer-transformerlm-librispeech 
& 2.1 & 5.4 & 11.9 & 7.4 & 38.9 & 33.0 & 30.6 & 23.5 & 44.9 & 79.5 & 44.5 & 75.4 & 17.8 & 17.0 
\\
hubert-large-ls960-ft 
& 2.0 & 4.1 & 8.4 & 5.4 & 29.6 & 22.8 & 20.8 & 16.0 & 32.0 & 60.0 & 33.7 & 59.1 & 14.4 & 10.9 
\\
hubert-xlarge-ls960-ft 
& 1.9 & 3.5 & 8.3 & 5.4 & 29.3 & 22.2 & 19.8 & 14.8 & 31.5 & 58.5 & 33.3 & 58.9 & 14.2 & 10.5 
\\
s2t-large-librispeech-asr 
& 3.3 & 8.1 & 14.9 & 9.4 & 54.5 & 40.3 & 38.1 & 30.7 & 50.2 & 79.2 & 53.4 & 79.5 & 21.6 & 18.0 
\\
s2t-medium-librispeech-asr 
& 3.6 & 8.2 & 15.7 & 9.7 & 58.1 & 42.4 & 39.3 & 31.3 & 52.6 & 79.8 & 60.3 & 85.3 & 22.9 & 19.7 
\\
stt\_en\_conformer\_ctc\_large 
& 2.1 & 4.2 & 4.4 & 2.1 & 11.3 & 8.2 & 7.4 & 4.0 & 13.5 & 30.5 & 15.9 & 39.9 & 6.7 & 8.2 
\\
stt\_en\_conformer\_transducer\_xlarge 
& 1.5 & 2.8 & 4.3 & 1.2 & 12.0 & 7.4 & 4.3 & 1.5 & 19.9 & 36.8 & 20.5 & 48.6 & 6.0 & 6.3 
\\
unispeech-sat-base-100h-libri-ft 
& 5.7 & 13.8 & 17.7 & 13.6 & 46.5 & 40.0 & 45.3 & 38.6 & 44.7 & 74.8 & 47.8 & 77.7 & 29.8 & 22.4 
\\
\bottomrule
\end{tabular}
\caption{English transcription WER (\%) with greedy decoding}\label{tab:english-asr-greedy}
\end{table}

\subsubsection{Beam search with temperature fallback}
\begin{table}[H]\centering
\scriptsize
\begin{tabular}{l|rrrrrrrrrrrrrr}
\toprule
\setlength{\tabcolsep}{2pt}\renewcommand{\arraystretch}{1.2}
Model
& \multicolumn{1}{c}{\rotatebox[origin=rc]{270}{LibriSpeech.test-clean}}
& \multicolumn{1}{c}{\rotatebox[origin=rc]{270}{LibriSpeech.test-other}}
& \multicolumn{1}{c}{\rotatebox[origin=rc]{270}{TED-LIUM3}}
& \multicolumn{1}{c}{\rotatebox[origin=rc]{270}{WSJ}}
& \multicolumn{1}{c}{\rotatebox[origin=rc]{270}{CallHome}}
& \multicolumn{1}{c}{\rotatebox[origin=rc]{270}{Switchboard}}
& \multicolumn{1}{c}{\rotatebox[origin=rc]{270}{CommonVoice5.1}}
& \multicolumn{1}{c}{\rotatebox[origin=rc]{270}{Artie}}
& \multicolumn{1}{c}{\rotatebox[origin=rc]{270}{CORAAL}}
& \multicolumn{1}{c}{\rotatebox[origin=rc]{270}{CHiME6}}
& \multicolumn{1}{c}{\rotatebox[origin=rc]{270}{AMI-IHM}}
& \multicolumn{1}{c}{\rotatebox[origin=rc]{270}{AMI-SDM1}}
& \multicolumn{1}{c}{\rotatebox[origin=rc]{270}{VoxPopuli.en}}
& \multicolumn{1}{c}{\rotatebox[origin=rc]{270}{Fleurs.en\_us}}
\\ \midrule
Whisper tiny.en 
& 5.4 & 12.8 & 5.4 & 4.6 & 21.4 & 16.0 & 23.5 & 18.4 & 21.4 & 42.0 & 22.7 & 54.2 & 10.9 & 10.0 
\\
Whisper tiny 
& 6.7 & 15.0 & 6.3 & 5.9 & 24.8 & 18.3 & 26.1 & 20.8 & 25.1 & 48.0 & 25.6 & 57.3 & 11.6 & 12.4 
\\
Whisper base.en 
& 4.1 & 9.6 & 4.6 & 4.0 & 18.3 & 14.2 & 17.5 & 13.2 & 18.5 & 35.2 & 21.1 & 49.0 & 9.3 & 7.1 
\\
Whisper base 
& 4.9 & 11.0 & 5.0 & 4.4 & 20.5 & 15.6 & 19.4 & 15.3 & 20.5 & 40.0 & 21.5 & 50.0 & 9.5 & 8.9 
\\
Whisper small.en 
& 3.2 & 6.7 & 4.3 & 3.0 & 17.2 & 13.4 & 12.6 & 9.2 & 17.5 & 29.5 & 17.9 & 42.5 & 8.1 & 5.3 
\\
Whisper small 
& 3.3 & 7.2 & 4.3 & 3.9 & 17.1 & 13.3 & 12.8 & 9.3 & 16.4 & 30.9 & 19.2 & 43.5 & 8.2 & 6.1 
\\
Whisper medium.en 
& 3.0 & 5.7 & 4.3 & 2.8 & 14.7 & 12.4 & 10.3 & 7.4 & 15.3 & 27.0 & 17.1 & 39.4 & 7.8 & 4.5 
\\
Whisper medium 
& 2.7 & 5.6 & 4.0 & 2.7 & 15.3 & 13.2 & 9.7 & 6.7 & 14.9 & 27.6 & 17.6 & 43.0 & 7.6 & 4.4 
\\
Whisper large 
& 2.8 & 5.7 & 4.3 & 3.5 & 16.2 & 14.2 & 8.9 & 6.4 & 15.1 & 25.2 & 17.6 & 37.1 & 7.2 & 4.5 
\\
Whisper large-v2
& 2.5 & 4.9 & 3.7 & 2.6 & 16.4 & 13.6 & 8.2 & 5.7 & 14.2 & 24.9 & 17.4 & 39.9 & 7.0 & 4.2
\\
\bottomrule
\end{tabular}
\caption{English transcription WER (\%) with beam search and temperature fallback}\label{tab:english-asr-bs5fb}
\end{table}

\subsection{Multilingual Transcription}

\subsubsection{Multilingual LibriSpeech}
\begin{table}[H]\centering
\scriptsize
\vskip 0.2em
\begin{tabular}{l|rrrrrrrr}
\toprule
\setlength{\tabcolsep}{-5pt}\renewcommand{\arraystretch}{1.2}
Model
& \multicolumn{1}{c}{\rotatebox[origin=rc]{270}{Dutch}}
& \multicolumn{1}{c}{\rotatebox[origin=rc]{270}{English}}
& \multicolumn{1}{c}{\rotatebox[origin=rc]{270}{French}}
& \multicolumn{1}{c}{\rotatebox[origin=rc]{270}{German}}
& \multicolumn{1}{c}{\rotatebox[origin=rc]{270}{Italian}}
& \multicolumn{1}{c}{\rotatebox[origin=rc]{270}{Polish}}
& \multicolumn{1}{c}{\rotatebox[origin=rc]{270}{Portuguese}}
& \multicolumn{1}{c}{\rotatebox[origin=rc]{270}{Spanish}}
\\ \midrule
Whisper tiny 
& 39.4 & 15.7 & 36.8 & 24.9 & 41.7 & 34.2 & 31.3 & 19.2 
\\
Whisper base 
& 28.4 & 11.7 & 26.6 & 17.7 & 31.1 & 22.8 & 21.9 & 12.8 
\\
Whisper small 
& 17.2 & 8.3 & 16.2 & 10.5 & 21.4 & 11.2 & 13.0 & 7.8 
\\
Whisper medium 
& 11.7 & 6.8 & 8.9 & 7.4 & 16.0 & 6.5 & 9.0 & 5.3 
\\
Whisper large 
& 10.2 & 6.3 & 8.9 & 6.6 & 14.3 & 6.6 & 9.2 & 5.4 
\\
Whisper large-v2 
& 9.3 & 6.2 & 7.3 & 5.5 & 13.8 & 5.0 & 6.8 & 4.2 
\\
\bottomrule
\end{tabular}
\caption{WER (\%) on MLS}\label{tab:mls-asr}
\end{table}

\vspace{-1em}

\subsubsection{Common Voice 9}
\begin{table}[H]\centering
\scriptsize
\vskip 0.2em
\begin{tabular}{l|rrrrrrrrrrrrr}
\toprule
\setlength{\tabcolsep}{-5pt}\renewcommand{\arraystretch}{1.2}
Model
& \multicolumn{1}{c}{\rotatebox[origin=rc]{270}{Arabic}}
& \multicolumn{1}{c}{\rotatebox[origin=rc]{270}{Bulgarian}}
& \multicolumn{1}{c}{\rotatebox[origin=rc]{270}{Bengali}}
& \multicolumn{1}{c}{\rotatebox[origin=rc]{270}{Catalan}}
& \multicolumn{1}{c}{\rotatebox[origin=rc]{270}{Czech}}
& \multicolumn{1}{c}{\rotatebox[origin=rc]{270}{Welsh}}
& \multicolumn{1}{c}{\rotatebox[origin=rc]{270}{Danish}}
& \multicolumn{1}{c}{\rotatebox[origin=rc]{270}{German}}
& \multicolumn{1}{c}{\rotatebox[origin=rc]{270}{Greek}}
& \multicolumn{1}{c}{\rotatebox[origin=rc]{270}{English}}
& \multicolumn{1}{c}{\rotatebox[origin=rc]{270}{Spanish}}
& \multicolumn{1}{c}{\rotatebox[origin=rc]{270}{Estonian}}
& \multicolumn{1}{c}{\rotatebox[origin=rc]{270}{Persian}}
\\ \midrule
Whisper tiny 
& 90.9 & 79.3 & 104.1 & 51.0 & 79.7 & 101.8 & 77.2 & 34.5 & 61.9 & 28.8 & 30.3 & 102.1 & 120.3 
\\
Whisper base 
& 84.4 & 68.1 & 103.7 & 39.9 & 63.1 & 93.8 & 57.5 & 24.5 & 51.5 & 21.9 & 19.6 & 88.1 & 99.0 
\\
Whisper small 
& 66.4 & 44.8 & 118.6 & 23.8 & 34.1 & 65.4 & 32.1 & 13.0 & 31.7 & 14.5 & 10.3 & 67.2 & 71.9 
\\
Whisper medium 
& 60.3 & 26.7 & 124.7 & 16.4 & 18.8 & 43.6 & 19.3 & 8.5 & 20.0 & 11.2 & 6.9 & 45.6 & 49.9 
\\
Whisper large 
& 56.0 & 24.1 & 106.0 & 15.3 & 17.1 & 40.3 & 18.3 & 7.7 & 18.3 & 10.1 & 6.4 & 41.4 & 44.8 
\\
Whisper large-v2 
& 53.8 & 19.9 & 103.4 & 14.1 & 13.5 & 34.2 & 14.4 & 6.4 & 16.0 & 9.4 & 5.6 & 35.1 & 39.4 
\\
\bottomrule
\end{tabular}
\vskip 0.2em
\begin{tabular}{l|rrrrrrrrrrrrr}
\toprule
\setlength{\tabcolsep}{-5pt}\renewcommand{\arraystretch}{1.2}
Model
& \multicolumn{1}{c}{\rotatebox[origin=rc]{270}{Finnish}}
& \multicolumn{1}{c}{\rotatebox[origin=rc]{270}{French}}
& \multicolumn{1}{c}{\rotatebox[origin=rc]{270}{Hindi}}
& \multicolumn{1}{c}{\rotatebox[origin=rc]{270}{Hungarian}}
& \multicolumn{1}{c}{\rotatebox[origin=rc]{270}{Indonesian}}
& \multicolumn{1}{c}{\rotatebox[origin=rc]{270}{Italian}}
& \multicolumn{1}{c}{\rotatebox[origin=rc]{270}{Japanese}}
& \multicolumn{1}{c}{\rotatebox[origin=rc]{270}{Lithuanian}}
& \multicolumn{1}{c}{\rotatebox[origin=rc]{270}{Latvian}}
& \multicolumn{1}{c}{\rotatebox[origin=rc]{270}{Malayalam}}
& \multicolumn{1}{c}{\rotatebox[origin=rc]{270}{Mongolian}}
& \multicolumn{1}{c}{\rotatebox[origin=rc]{270}{Dutch}}
& \multicolumn{1}{c}{\rotatebox[origin=rc]{270}{Polish}}
\\ \midrule
Whisper tiny 
& 68.5 & 49.7 & 108.3 & 87.0 & 49.6 & 44.5 & 36.1 & 103.5 & 87.8 & 102.7 & 123.0 & 43.6 & 45.3 
\\
Whisper base 
& 52.9 & 37.3 & 106.5 & 71.9 & 36.1 & 30.5 & 24.2 & 91.3 & 78.0 & 122.9 & 137.0 & 29.5 & 32.8 
\\
Whisper small 
& 30.5 & 22.7 & 43.6 & 44.4 & 18.4 & 16.0 & 14.0 & 72.8 & 54.6 & 104.8 & 225.8 & 14.2 & 16.9 
\\
Whisper medium 
& 18.8 & 16.0 & 31.5 & 26.9 & 11.6 & 9.4 & 10.5 & 49.4 & 37.2 & 137.8 & 113.4 & 8.0 & 10.1 
\\
Whisper large 
& 17.0 & 14.7 & 25.0 & 23.5 & 10.6 & 8.1 & 9.4 & 43.9 & 34.8 & 107.1 & 117.4 & 7.1 & 9.0 
\\
Whisper large-v2 
& 14.4 & 13.9 & 21.9 & 19.7 & 8.5 & 7.1 & 9.1 & 35.2 & 25.5 & 103.2 & 128.4 & 5.8 & 7.6 
\\
\bottomrule
\end{tabular}
\vskip 0.2em
\begin{tabular}{l|rrrrrrrrrrrrr}
\toprule
\setlength{\tabcolsep}{-5pt}\renewcommand{\arraystretch}{1.2}
Model
& \multicolumn{1}{c}{\rotatebox[origin=rc]{270}{Portuguese}}
& \multicolumn{1}{c}{\rotatebox[origin=rc]{270}{Romanian}}
& \multicolumn{1}{c}{\rotatebox[origin=rc]{270}{Russian}}
& \multicolumn{1}{c}{\rotatebox[origin=rc]{270}{Slovak}}
& \multicolumn{1}{c}{\rotatebox[origin=rc]{270}{Slovenian}}
& \multicolumn{1}{c}{\rotatebox[origin=rc]{270}{Serbian}}
& \multicolumn{1}{c}{\rotatebox[origin=rc]{270}{Swedish}}
& \multicolumn{1}{c}{\rotatebox[origin=rc]{270}{Tamil}}
& \multicolumn{1}{c}{\rotatebox[origin=rc]{270}{Thai}}
& \multicolumn{1}{c}{\rotatebox[origin=rc]{270}{Turkish}}
& \multicolumn{1}{c}{\rotatebox[origin=rc]{270}{Urdu}}
& \multicolumn{1}{c}{\rotatebox[origin=rc]{270}{Vietnamese}}
& \multicolumn{1}{c}{\rotatebox[origin=rc]{270}{Chinese}}
\\ \midrule
Whisper tiny 
& 35.2 & 68.2 & 40.6 & 104.0 & 82.0 & 106.1 & 58.2 & 105.7 & 55.9 & 53.6 & 74.7 & 69.3 & 52.4 
\\
Whisper base 
& 23.7 & 55.9 & 28.8 & 87.2 & 70.3 & 103.0 & 42.4 & 49.5 & 32.1 & 38.6 & 58.6 & 51.6 & 44.9 
\\
Whisper small 
& 12.5 & 33.2 & 15.0 & 60.4 & 45.5 & 101.3 & 22.1 & 28.7 & 18.1 & 23.7 & 39.1 & 33.3 & 29.4 
\\
Whisper medium 
& 8.1 & 21.5 & 9.3 & 42.0 & 29.8 & 85.6 & 13.7 & 19.6 & 10.5 & 17.7 & 29.9 & 24.4 & 23.2 
\\
Whisper large 
& 7.1 & 19.8 & 8.2 & 37.9 & 25.1 & 87.4 & 12.4 & 17.6 & 8.8 & 16.6 & 28.1 & 19.9 & 29.1 
\\
Whisper large-v2 
& 6.3 & 15.8 & 7.1 & 31.9 & 20.6 & 70.5 & 10.6 & 16.1 & 8.0 & 14.5 & 24.2 & 18.2 & 26.8 
\\
\bottomrule
\end{tabular}
\caption{WER (\%) on CommonVoice9}\label{tab:cv9-asr}
\end{table}

\vspace{-1em}

\subsubsection{VoxPopuli}
\begin{table}[H]\centering
\scriptsize
\vskip 0.2em
\begin{tabular}{l|rrrrrrrrrrrrrrrrr}
\toprule
\setlength{\tabcolsep}{-5pt}\renewcommand{\arraystretch}{1.2}
Model
& \multicolumn{1}{c}{\rotatebox[origin=rc]{270}{Czech}}
& \multicolumn{1}{c}{\rotatebox[origin=rc]{270}{German}}
& \multicolumn{1}{c}{\rotatebox[origin=rc]{270}{English}}
& \multicolumn{1}{c}{\rotatebox[origin=rc]{270}{\texttt{en\_accented}}}
& \multicolumn{1}{c}{\rotatebox[origin=rc]{270}{Spanish}}
& \multicolumn{1}{c}{\rotatebox[origin=rc]{270}{Estonian}}
& \multicolumn{1}{c}{\rotatebox[origin=rc]{270}{Finnish}}
& \multicolumn{1}{c}{\rotatebox[origin=rc]{270}{French}}
& \multicolumn{1}{c}{\rotatebox[origin=rc]{270}{Croatian}}
& \multicolumn{1}{c}{\rotatebox[origin=rc]{270}{Hungarian}}
& \multicolumn{1}{c}{\rotatebox[origin=rc]{270}{Italian}}
& \multicolumn{1}{c}{\rotatebox[origin=rc]{270}{Lithuanian}}
& \multicolumn{1}{c}{\rotatebox[origin=rc]{270}{Dutch}}
& \multicolumn{1}{c}{\rotatebox[origin=rc]{270}{Polish}}
& \multicolumn{1}{c}{\rotatebox[origin=rc]{270}{Romanian}}
& \multicolumn{1}{c}{\rotatebox[origin=rc]{270}{Slovak}}
& \multicolumn{1}{c}{\rotatebox[origin=rc]{270}{Slovenian}}
\\ \midrule
Whisper tiny 
& 73.5 & 27.4 & 11.6 & 18.8 & 19.7 & 99.2 & 54.1 & 32.9 & 72.4 & 74.5 & 40.5 & 93.1 & 41.9 & 31.4 & 65.9 & 78.7 & 81.9 
\\
Whisper base 
& 54.7 & 20.6 & 9.5 & 17.5 & 14.4 & 83.0 & 39.7 & 24.9 & 53.6 & 52.6 & 30.8 & 82.1 & 29.4 & 22.1 & 49.3 & 63.7 & 70.5 
\\
Whisper small 
& 28.8 & 14.8 & 8.2 & 19.2 & 11.1 & 59.2 & 24.9 & 15.7 & 33.7 & 31.3 & 22.9 & 60.1 & 18.8 & 13.3 & 28.6 & 37.3 & 50.8 
\\
Whisper medium 
& 18.4 & 12.4 & 7.6 & 19.1 & 9.6 & 38.2 & 16.6 & 12.2 & 23.9 & 19.3 & 19.7 & 39.3 & 14.9 & 10.1 & 18.4 & 23.0 & 36.3 
\\
Whisper large 
& 15.9 & 11.9 & 7.2 & 20.8 & 8.8 & 33.3 & 15.5 & 11.0 & 19.0 & 16.8 & 18.4 & 35.0 & 14.0 & 9.0 & 17.0 & 19.1 & 31.3 
\\
Whisper large-v2 
& 12.6 & 11.2 & 7.0 & 18.6 & 8.2 & 28.7 & 12.4 & 11.4 & 16.1 & 13.8 & 19.0 & 33.2 & 12.9 & 7.8 & 14.4 & 15.4 & 27.9 
\\
\bottomrule
\end{tabular}
\caption{WER (\%) on VoxPopuli}\label{tab:voxpopuli-asr}
\end{table}

\vspace{-1em}

\subsubsection{Fleurs}
\begin{table}[H]\centering
\scriptsize
\vskip 0.2em
\begin{tabular}{l|rrrrrrrrrrrrrr}
\toprule
\setlength{\tabcolsep}{-5pt}\renewcommand{\arraystretch}{1.2}
Model
& \multicolumn{1}{c}{\rotatebox[origin=rc]{270}{Afrikaans}}
& \multicolumn{1}{c}{\rotatebox[origin=rc]{270}{Amharic}}
& \multicolumn{1}{c}{\rotatebox[origin=rc]{270}{Arabic}}
& \multicolumn{1}{c}{\rotatebox[origin=rc]{270}{Assamese}}
& \multicolumn{1}{c}{\rotatebox[origin=rc]{270}{Azerbaijani}}
& \multicolumn{1}{c}{\rotatebox[origin=rc]{270}{Belarusian}}
& \multicolumn{1}{c}{\rotatebox[origin=rc]{270}{Bulgarian}}
& \multicolumn{1}{c}{\rotatebox[origin=rc]{270}{Bengali}}
& \multicolumn{1}{c}{\rotatebox[origin=rc]{270}{Bosnian}}
& \multicolumn{1}{c}{\rotatebox[origin=rc]{270}{Catalan}}
& \multicolumn{1}{c}{\rotatebox[origin=rc]{270}{Chinese}}
& \multicolumn{1}{c}{\rotatebox[origin=rc]{270}{Czech}}
& \multicolumn{1}{c}{\rotatebox[origin=rc]{270}{Welsh}}
& \multicolumn{1}{c}{\rotatebox[origin=rc]{270}{Danish}}
\\ \midrule
Whisper tiny 
& 91.2 & 122.9 & 63.4 & 102.0 & 93.1 & 94.0 & 81.0 & 101.6 & 82.1 & 42.8 & 40.5 & 82.8 & 101.3 & 82.0 
\\
Whisper base 
& 81.5 & 196.8 & 48.8 & 102.0 & 76.4 & 91.3 & 65.1 & 100.6 & 66.7 & 29.0 & 34.1 & 66.0 & 85.3 & 57.6 
\\
Whisper small 
& 61.1 & 120.2 & 30.6 & 108.0 & 49.1 & 75.1 & 37.3 & 104.4 & 39.4 & 16.2 & 20.8 & 37.6 & 59.3 & 32.8 
\\
Whisper medium 
& 44.9 & 229.3 & 20.4 & 102.3 & 33.1 & 60.4 & 21.4 & 100.6 & 23.9 & 9.6 & 12.1 & 21.3 & 40.8 & 19.5 
\\
Whisper large 
& 42.6 & 129.3 & 18.1 & 105.6 & 28.7 & 56.6 & 18.4 & 104.9 & 20.7 & 8.0 & 19.6 & 17.4 & 36.6 & 16.8 
\\
Whisper large-v2 
& 36.7 & 140.3 & 16.0 & 106.2 & 23.4 & 45.4 & 14.6 & 104.1 & 15.7 & 7.3 & 14.7 & 13.3 & 33.0 & 13.8 
\\
\bottomrule
\end{tabular}
\vskip 0.2em
\begin{tabular}{l|rrrrrrrrrrrrrr}
\toprule
\setlength{\tabcolsep}{-5pt}\renewcommand{\arraystretch}{1.2}
Model
& \multicolumn{1}{c}{\rotatebox[origin=rc]{270}{German}}
& \multicolumn{1}{c}{\rotatebox[origin=rc]{270}{Greek}}
& \multicolumn{1}{c}{\rotatebox[origin=rc]{270}{English}}
& \multicolumn{1}{c}{\rotatebox[origin=rc]{270}{Spanish}}
& \multicolumn{1}{c}{\rotatebox[origin=rc]{270}{Estonian}}
& \multicolumn{1}{c}{\rotatebox[origin=rc]{270}{Persian}}
& \multicolumn{1}{c}{\rotatebox[origin=rc]{270}{Finnish}}
& \multicolumn{1}{c}{\rotatebox[origin=rc]{270}{Tagalog}}
& \multicolumn{1}{c}{\rotatebox[origin=rc]{270}{French}}
& \multicolumn{1}{c}{\rotatebox[origin=rc]{270}{Galician}}
& \multicolumn{1}{c}{\rotatebox[origin=rc]{270}{Gujarati}}
& \multicolumn{1}{c}{\rotatebox[origin=rc]{270}{Hausa}}
& \multicolumn{1}{c}{\rotatebox[origin=rc]{270}{Hebrew}}
& \multicolumn{1}{c}{\rotatebox[origin=rc]{270}{Hindi}}
\\ \midrule
Whisper tiny 
& 27.8 & 67.4 & 12.4 & 15.9 & 94.8 & 101.8 & 59.5 & 65.6 & 41.4 & 54.8 & 101.2 & 100.2 & 71.6 & 102.3 
\\
Whisper base 
& 17.9 & 53.5 & 8.9 & 9.9 & 77.9 & 86.1 & 43.1 & 45.8 & 28.5 & 47.4 & 101.4 & 98.6 & 61.7 & 101.1 
\\
Whisper small 
& 10.2 & 30.8 & 6.1 & 5.6 & 51.3 & 55.8 & 24.0 & 27.7 & 15.0 & 30.2 & 106.4 & 90.1 & 44.4 & 38.4 
\\
Whisper medium 
& 6.5 & 19.0 & 4.4 & 3.6 & 29.8 & 41.0 & 13.9 & 19.1 & 8.7 & 21.2 & 104.8 & 106.6 & 33.1 & 26.8 
\\
Whisper large 
& 5.5 & 18.7 & 4.5 & 3.5 & 25.5 & 36.1 & 12.2 & 15.8 & 7.7 & 19.0 & 103.9 & 87.0 & 30.2 & 26.9 
\\
Whisper large-v2 
& 4.5 & 12.5 & 4.2 & 3.0 & 21.9 & 32.9 & 9.7 & 13.8 & 8.3 & 15.4 & 102.7 & 88.9 & 27.1 & 21.5 
\\
\bottomrule
\end{tabular}
\vskip 0.2em
\begin{tabular}{l|rrrrrrrrrrrrrr}
\toprule
\setlength{\tabcolsep}{-5pt}\renewcommand{\arraystretch}{1.2}
Model
& \multicolumn{1}{c}{\rotatebox[origin=rc]{270}{Croatian}}
& \multicolumn{1}{c}{\rotatebox[origin=rc]{270}{Hungarian}}
& \multicolumn{1}{c}{\rotatebox[origin=rc]{270}{Armenian}}
& \multicolumn{1}{c}{\rotatebox[origin=rc]{270}{Indonesian}}
& \multicolumn{1}{c}{\rotatebox[origin=rc]{270}{Icelandic}}
& \multicolumn{1}{c}{\rotatebox[origin=rc]{270}{Italian}}
& \multicolumn{1}{c}{\rotatebox[origin=rc]{270}{Japanese}}
& \multicolumn{1}{c}{\rotatebox[origin=rc]{270}{Javanese}}
& \multicolumn{1}{c}{\rotatebox[origin=rc]{270}{Georgian}}
& \multicolumn{1}{c}{\rotatebox[origin=rc]{270}{Kazakh}}
& \multicolumn{1}{c}{\rotatebox[origin=rc]{270}{Khmer}}
& \multicolumn{1}{c}{\rotatebox[origin=rc]{270}{Kannada}}
& \multicolumn{1}{c}{\rotatebox[origin=rc]{270}{Korean}}
& \multicolumn{1}{c}{\rotatebox[origin=rc]{270}{Luxembourgish}}
\\ \midrule
Whisper tiny 
& 79.0 & 83.8 & 118.6 & 51.7 & 113.3 & 29.8 & 37.0 & 107.3 & 123.0 & 165.2 & 100.6 & 100.7 & 36.1 & 99.1 
\\
Whisper base 
& 59.1 & 65.0 & 126.3 & 33.1 & 95.5 & 17.9 & 22.8 & 89.5 & 114.7 & 109.2 & 101.6 & 107.2 & 27.8 & 100.7 
\\
Whisper small 
& 33.4 & 38.9 & 86.6 & 16.3 & 72.6 & 9.8 & 12.0 & 88.6 & 118.3 & 70.3 & 104.4 & 100.4 & 19.6 & 100.1 
\\
Whisper medium 
& 19.3 & 24.3 & 60.1 & 10.2 & 49.9 & 5.2 & 7.1 & 67.9 & 117.3 & 48.8 & 98.9 & 77.7 & 16.4 & 90.0 
\\
Whisper large 
& 16.7 & 21.0 & 53.7 & 8.5 & 43.0 & 4.2 & 6.4 & 87.0 & 100.5 & 43.8 & 96.0 & 69.8 & 15.2 & 86.5 
\\
Whisper large-v2 
& 13.4 & 17.0 & 44.6 & 7.1 & 38.2 & 4.0 & 5.3 & nan & 105.0 & 37.7 & 99.7 & 37.0 & 14.3 & 88.0 
\\
\bottomrule
\end{tabular}
\vskip 0.2em
\begin{tabular}{l|rrrrrrrrrrrrrr}
\toprule
\setlength{\tabcolsep}{-5pt}\renewcommand{\arraystretch}{1.2}
Model
& \multicolumn{1}{c}{\rotatebox[origin=rc]{270}{Lingala}}
& \multicolumn{1}{c}{\rotatebox[origin=rc]{270}{Lao}}
& \multicolumn{1}{c}{\rotatebox[origin=rc]{270}{Lithuanian}}
& \multicolumn{1}{c}{\rotatebox[origin=rc]{270}{Latvian}}
& \multicolumn{1}{c}{\rotatebox[origin=rc]{270}{Maori}}
& \multicolumn{1}{c}{\rotatebox[origin=rc]{270}{Macedonian}}
& \multicolumn{1}{c}{\rotatebox[origin=rc]{270}{Malayalam}}
& \multicolumn{1}{c}{\rotatebox[origin=rc]{270}{Mongolian}}
& \multicolumn{1}{c}{\rotatebox[origin=rc]{270}{Marathi}}
& \multicolumn{1}{c}{\rotatebox[origin=rc]{270}{Malay}}
& \multicolumn{1}{c}{\rotatebox[origin=rc]{270}{Maltese}}
& \multicolumn{1}{c}{\rotatebox[origin=rc]{270}{Myanmar}}
& \multicolumn{1}{c}{\rotatebox[origin=rc]{270}{Norwegian}}
& \multicolumn{1}{c}{\rotatebox[origin=rc]{270}{Nepali}}
\\ \midrule
Whisper tiny 
& 105.4 & 115.1 & 98.5 & 91.6 & 94.5 & 73.3 & 101.5 & 113.7 & 100.3 & 51.2 & 100.8 & 124.8 & 62.0 & 101.8 
\\
Whisper base 
& 96.7 & 105.1 & 87.3 & 79.8 & 77.5 & 59.9 & 107.4 & 125.7 & 100.3 & 35.1 & 97.6 & 122.6 & 44.0 & 102.4 
\\
Whisper small 
& 91.3 & 102.2 & 65.6 & 53.2 & 59.5 & 36.9 & 100.9 & 144.2 & 60.2 & 18.9 & 92.2 & 110.1 & 24.2 & 69.5 
\\
Whisper medium 
& 83.2 & 101.4 & 41.1 & 32.0 & 77.8 & 22.0 & 101.1 & 103.7 & 63.2 & 12.2 & 83.2 & 123.0 & 12.9 & 54.4 
\\
Whisper large 
& 76.8 & 101.6 & 35.2 & 28.3 & 45.7 & 20.6 & 101.4 & 106.2 & 43.7 & 10.2 & 80.5 & 124.5 & 11.4 & 52.2 
\\
Whisper large-v2 
& 75.6 & 101.5 & 28.1 & 23.1 & 38.5 & 16.5 & 100.7 & 110.5 & 38.3 & 8.7 & 76.6 & 115.7 & 9.5 & 47.1 
\\
\bottomrule
\end{tabular}
\vskip 0.2em
\begin{tabular}{l|rrrrrrrrrrrrrr}
\toprule
\setlength{\tabcolsep}{-5pt}\renewcommand{\arraystretch}{1.2}
Model
& \multicolumn{1}{c}{\rotatebox[origin=rc]{270}{Dutch}}
& \multicolumn{1}{c}{\rotatebox[origin=rc]{270}{Occitan}}
& \multicolumn{1}{c}{\rotatebox[origin=rc]{270}{Punjabi}}
& \multicolumn{1}{c}{\rotatebox[origin=rc]{270}{Polish}}
& \multicolumn{1}{c}{\rotatebox[origin=rc]{270}{Pashto}}
& \multicolumn{1}{c}{\rotatebox[origin=rc]{270}{Portuguese}}
& \multicolumn{1}{c}{\rotatebox[origin=rc]{270}{Romanian}}
& \multicolumn{1}{c}{\rotatebox[origin=rc]{270}{Russian}}
& \multicolumn{1}{c}{\rotatebox[origin=rc]{270}{Sindhi}}
& \multicolumn{1}{c}{\rotatebox[origin=rc]{270}{Slovak}}
& \multicolumn{1}{c}{\rotatebox[origin=rc]{270}{Slovenian}}
& \multicolumn{1}{c}{\rotatebox[origin=rc]{270}{Shona}}
& \multicolumn{1}{c}{\rotatebox[origin=rc]{270}{Somali}}
& \multicolumn{1}{c}{\rotatebox[origin=rc]{270}{Serbian}}
\\ \midrule
Whisper tiny 
& 49.0 & 95.9 & 102.6 & 45.6 & 105.6 & 20.1 & 74.7 & 31.1 & 105.8 & 77.2 & 87.2 & 128.1 & 105.6 & 83.7 
\\
Whisper base 
& 33.0 & 82.9 & 101.5 & 30.8 & 99.0 & 13.0 & 56.0 & 20.5 & 103.9 & 60.6 & 74.6 & 126.0 & 109.6 & 64.3 
\\
Whisper small 
& 16.4 & 87.3 & 103.6 & 14.7 & 92.9 & 7.3 & 29.8 & 11.4 & 131.7 & 33.3 & 49.3 & 140.0 & 105.3 & 42.2 
\\
Whisper medium 
& 9.9 & 79.5 & 102.0 & 8.0 & 119.4 & 5.0 & 20.0 & 7.2 & 147.0 & 17.3 & 31.9 & 143.9 & 104.0 & 44.9 
\\
Whisper large 
& 8.3 & 75.9 & 102.8 & 7.2 & 92.7 & 4.8 & 15.4 & 6.4 & 177.9 & 15.7 & 27.8 & 130.0 & 103.5 & 29.2 
\\
Whisper large-v2 
& 6.7 & 75.3 & 102.4 & 5.4 & 93.7 & 4.3 & 14.4 & 5.6 & 156.5 & 11.7 & 23.1 & 121.0 & 102.9 & 33.9 
\\
\bottomrule
\end{tabular}
\vskip 0.2em
\begin{tabular}{l|rrrrrrrrrrrr}
\toprule
\setlength{\tabcolsep}{-5pt}\renewcommand{\arraystretch}{1.2}
Model
& \multicolumn{1}{c}{\rotatebox[origin=rc]{270}{Swedish}}
& \multicolumn{1}{c}{\rotatebox[origin=rc]{270}{Swahili}}
& \multicolumn{1}{c}{\rotatebox[origin=rc]{270}{Tamil}}
& \multicolumn{1}{c}{\rotatebox[origin=rc]{270}{Telugu}}
& \multicolumn{1}{c}{\rotatebox[origin=rc]{270}{Tajik}}
& \multicolumn{1}{c}{\rotatebox[origin=rc]{270}{Thai}}
& \multicolumn{1}{c}{\rotatebox[origin=rc]{270}{Turkish}}
& \multicolumn{1}{c}{\rotatebox[origin=rc]{270}{Ukrainian}}
& \multicolumn{1}{c}{\rotatebox[origin=rc]{270}{Urdu}}
& \multicolumn{1}{c}{\rotatebox[origin=rc]{270}{Uzbek}}
& \multicolumn{1}{c}{\rotatebox[origin=rc]{270}{Vietnamese}}
& \multicolumn{1}{c}{\rotatebox[origin=rc]{270}{Yoruba}}
\\ \midrule
Whisper tiny 
& 52.7 & 100.9 & 99.9 & 105.1 & 101.7 & 58.8 & 42.5 & 51.2 & 65.2 & 105.2 & 60.0 & 106.4 
\\
Whisper base 
& 37.4 & 92.5 & 58.7 & 105.2 & 109.3 & 38.2 & 27.5 & 37.7 & 52.0 & 114.0 & 40.5 & 101.8 
\\
Whisper small 
& 20.8 & 73.7 & 35.2 & 98.2 & 84.3 & 21.9 & 15.9 & 19.3 & 37.3 & 107.7 & 21.2 & 116.4 
\\
Whisper medium 
& 11.2 & 52.8 & 23.1 & 82.8 & 74.0 & 15.4 & 10.4 & 11.6 & 28.2 & 109.6 & 12.7 & 105.1 
\\
Whisper large 
& 10.5 & 47.9 & 20.6 & 100.6 & 74.5 & 13.2 & 9.4 & 10.3 & 25.0 & 93.3 & 10.7 & 111.7 
\\
Whisper large-v2 
& 8.5 & 39.3 & 17.5 & 99.0 & 85.8 & 11.5 & 8.4 & 8.6 & 22.6 & 90.2 & 10.3 & 94.8 
\\
\bottomrule
\end{tabular}
\caption{WER (\%) on Fleurs}\label{tab:fleurs-asr}
\end{table}

\subsection{Speech Translation}

\subsubsection{Fleurs}
\begin{table}[H]\centering
\scriptsize
\vskip 0.2em
\begin{tabular}{l|rrrrrrrrrrrrrr}
\toprule
\setlength{\tabcolsep}{-5pt}\renewcommand{\arraystretch}{1.2}
Model
& \multicolumn{1}{c}{\rotatebox[origin=rc]{270}{Afrikaans}}
& \multicolumn{1}{c}{\rotatebox[origin=rc]{270}{Amharic}}
& \multicolumn{1}{c}{\rotatebox[origin=rc]{270}{Arabic}}
& \multicolumn{1}{c}{\rotatebox[origin=rc]{270}{Assamese}}
& \multicolumn{1}{c}{\rotatebox[origin=rc]{270}{Azerbaijani}}
& \multicolumn{1}{c}{\rotatebox[origin=rc]{270}{Belarusian}}
& \multicolumn{1}{c}{\rotatebox[origin=rc]{270}{Bulgarian}}
& \multicolumn{1}{c}{\rotatebox[origin=rc]{270}{Bengali}}
& \multicolumn{1}{c}{\rotatebox[origin=rc]{270}{Bosnian}}
& \multicolumn{1}{c}{\rotatebox[origin=rc]{270}{Catalan}}
& \multicolumn{1}{c}{\rotatebox[origin=rc]{270}{Chinese}}
& \multicolumn{1}{c}{\rotatebox[origin=rc]{270}{Czech}}
& \multicolumn{1}{c}{\rotatebox[origin=rc]{270}{Welsh}}
& \multicolumn{1}{c}{\rotatebox[origin=rc]{270}{Danish}}
\\ \midrule
Whisper tiny 
& 1.6 & 0.1 & 0.1 & 0.4 & 0.1 & 0.8 & 0.4 & 0.4 & 0.4 & 5.2 & 0.6 & 0.6 & 0.6 & 0.7 
\\
Whisper base 
& 4.4 & 0.3 & 1.0 & 0.4 & 0.8 & 3.3 & 2.7 & 0.7 & 4.1 & 13.1 & 1.9 & 2.7 & 0.7 & 5.0 
\\
Whisper small 
& 18.1 & 0.2 & 10.6 & 1.2 & 5.8 & 7.1 & 14.8 & 2.7 & 16.8 & 25.1 & 9.3 & 14.2 & 1.3 & 18.1 
\\
Whisper medium 
& 29.5 & 0.9 & 19.9 & 3.5 & 11.7 & 9.8 & 23.9 & 10.6 & 26.0 & 31.9 & 15.1 & 23.6 & 8.4 & 28.6 
\\
Whisper large 
& 31.6 & 1.1 & 23.8 & 3.9 & 13.1 & 11.0 & 26.2 & 12.0 & 28.0 & 33.7 & 16.8 & 25.6 & 11.2 & 31.6 
\\
Whisper large-v2 
& 34.1 & 1.9 & 25.5 & 5.4 & 13.7 & 11.7 & 28.5 & 13.2 & 29.7 & 34.2 & 18.4 & 27.8 & 13.0 & 32.7 
\\
\bottomrule
\end{tabular}
\vskip 0.2em
\begin{tabular}{l|rrrrrrrrrrrrrr}
\toprule
\setlength{\tabcolsep}{-5pt}\renewcommand{\arraystretch}{1.2}
Model
& \multicolumn{1}{c}{\rotatebox[origin=rc]{270}{German}}
& \multicolumn{1}{c}{\rotatebox[origin=rc]{270}{Greek}}
& \multicolumn{1}{c}{\rotatebox[origin=rc]{270}{English}}
& \multicolumn{1}{c}{\rotatebox[origin=rc]{270}{Spanish}}
& \multicolumn{1}{c}{\rotatebox[origin=rc]{270}{Estonian}}
& \multicolumn{1}{c}{\rotatebox[origin=rc]{270}{Persian}}
& \multicolumn{1}{c}{\rotatebox[origin=rc]{270}{Finnish}}
& \multicolumn{1}{c}{\rotatebox[origin=rc]{270}{Tagalog}}
& \multicolumn{1}{c}{\rotatebox[origin=rc]{270}{French}}
& \multicolumn{1}{c}{\rotatebox[origin=rc]{270}{Galician}}
& \multicolumn{1}{c}{\rotatebox[origin=rc]{270}{Gujarati}}
& \multicolumn{1}{c}{\rotatebox[origin=rc]{270}{Hausa}}
& \multicolumn{1}{c}{\rotatebox[origin=rc]{270}{Hebrew}}
& \multicolumn{1}{c}{\rotatebox[origin=rc]{270}{Hindi}}
\\ \midrule
Whisper tiny 
& 5.2 & 0.1 & 68.6 & 7.7 & 0.1 & 0.1 & 0.2 & 0.8 & 4.7 & 4.0 & 0.7 & 0.1 & 0.2 & 1.0 
\\
Whisper base 
& 13.7 & 0.7 & 73.3 & 12.4 & 0.3 & 0.2 & 0.5 & 2.1 & 13.1 & 10.5 & 1.5 & 0.0 & 0.6 & 3.4 
\\
Whisper small 
& 25.9 & 11.6 & 77.3 & 18.2 & 3.6 & 5.8 & 7.3 & 12.0 & 23.5 & 17.5 & 3.9 & 0.3 & 5.4 & 11.1 
\\
Whisper medium 
& 31.4 & 19.9 & 79.2 & 21.4 & 13.5 & 15.0 & 18.5 & 20.5 & 28.6 & 24.7 & 12.8 & 0.5 & 15.9 & 19.4 
\\
Whisper large 
& 34.3 & 21.7 & 77.8 & 22.8 & 15.9 & 17.6 & 20.6 & 22.7 & 31.6 & 26.0 & 14.8 & 0.5 & 19.6 & 20.7 
\\
Whisper large-v2 
& 34.6 & 23.7 & 80.2 & 23.3 & 18.7 & 19.6 & 22.1 & 24.4 & 32.2 & 27.9 & 16.2 & 0.4 & 21.8 & 22.0 
\\
\bottomrule
\end{tabular}
\vskip 0.2em
\begin{tabular}{l|rrrrrrrrrrrrrr}
\toprule
\setlength{\tabcolsep}{-5pt}\renewcommand{\arraystretch}{1.2}
Model
& \multicolumn{1}{c}{\rotatebox[origin=rc]{270}{Croatian}}
& \multicolumn{1}{c}{\rotatebox[origin=rc]{270}{Hungarian}}
& \multicolumn{1}{c}{\rotatebox[origin=rc]{270}{Armenian}}
& \multicolumn{1}{c}{\rotatebox[origin=rc]{270}{Indonesian}}
& \multicolumn{1}{c}{\rotatebox[origin=rc]{270}{Icelandic}}
& \multicolumn{1}{c}{\rotatebox[origin=rc]{270}{Italian}}
& \multicolumn{1}{c}{\rotatebox[origin=rc]{270}{Japanese}}
& \multicolumn{1}{c}{\rotatebox[origin=rc]{270}{Javanese}}
& \multicolumn{1}{c}{\rotatebox[origin=rc]{270}{Georgian}}
& \multicolumn{1}{c}{\rotatebox[origin=rc]{270}{Kazakh}}
& \multicolumn{1}{c}{\rotatebox[origin=rc]{270}{Khmer}}
& \multicolumn{1}{c}{\rotatebox[origin=rc]{270}{Kannada}}
& \multicolumn{1}{c}{\rotatebox[origin=rc]{270}{Korean}}
& \multicolumn{1}{c}{\rotatebox[origin=rc]{270}{Luxembourgish}}
\\ \midrule
Whisper tiny 
& 0.6 & 0.1 & 0.1 & 0.3 & 0.4 & 5.3 & 0.2 & 0.2 & 0.1 & 0.1 & 0.1 & 0.8 & 0.5 & 0.8 
\\
Whisper base 
& 3.7 & 0.2 & 0.1 & 2.6 & 0.4 & 11.3 & 1.5 & 0.2 & 0.2 & 0.2 & 0.1 & 0.9 & 3.7 & 1.7 
\\
Whisper small 
& 14.6 & 4.8 & 0.7 & 16.4 & 1.8 & 17.8 & 9.6 & 1.4 & 0.2 & 0.8 & 0.5 & 2.3 & 12.2 & 5.7 
\\
Whisper medium 
& 23.0 & 15.5 & 10.4 & 24.1 & 6.8 & 21.6 & 14.9 & 5.0 & 1.3 & 4.3 & 3.3 & 8.5 & 19.2 & 13.6 
\\
Whisper large 
& 25.4 & 18.3 & 13.2 & 27.2 & 6.6 & 23.5 & 17.0 & 5.1 & 2.7 & 6.3 & 5.2 & 9.9 & 20.0 & 15.4 
\\
Whisper large-v2 
& 27.0 & 21.2 & 16.0 & 29.1 & 9.1 & 23.6 & 18.9 & 6.2 & 2.4 & 5.4 & 6.1 & 11.6 & 21.3 & 16.8 
\\
\bottomrule
\end{tabular}
\vskip 0.2em
\begin{tabular}{l|rrrrrrrrrrrrrr}
\toprule
\setlength{\tabcolsep}{-5pt}\renewcommand{\arraystretch}{1.2}
Model
& \multicolumn{1}{c}{\rotatebox[origin=rc]{270}{Lingala}}
& \multicolumn{1}{c}{\rotatebox[origin=rc]{270}{Lao}}
& \multicolumn{1}{c}{\rotatebox[origin=rc]{270}{Lithuanian}}
& \multicolumn{1}{c}{\rotatebox[origin=rc]{270}{Latvian}}
& \multicolumn{1}{c}{\rotatebox[origin=rc]{270}{Maori}}
& \multicolumn{1}{c}{\rotatebox[origin=rc]{270}{Macedonian}}
& \multicolumn{1}{c}{\rotatebox[origin=rc]{270}{Malayalam}}
& \multicolumn{1}{c}{\rotatebox[origin=rc]{270}{Mongolian}}
& \multicolumn{1}{c}{\rotatebox[origin=rc]{270}{Marathi}}
& \multicolumn{1}{c}{\rotatebox[origin=rc]{270}{Malay}}
& \multicolumn{1}{c}{\rotatebox[origin=rc]{270}{Maltese}}
& \multicolumn{1}{c}{\rotatebox[origin=rc]{270}{Myanmar}}
& \multicolumn{1}{c}{\rotatebox[origin=rc]{270}{Norwegian}}
& \multicolumn{1}{c}{\rotatebox[origin=rc]{270}{Nepali}}
\\ \midrule
Whisper tiny 
& 0.1 & 0.2 & 0.1 & 0.2 & 0.3 & 1.0 & 0.8 & 0.1 & 0.2 & 0.3 & 0.6 & 0.1 & 1.4 & 0.1 
\\
Whisper base 
& 0.1 & 0.3 & 0.3 & 0.4 & 1.0 & 5.4 & 1.4 & 0.1 & 0.9 & 2.1 & 1.4 & 0.1 & 8.4 & 0.3 
\\
Whisper small 
& 0.5 & 2.0 & 1.9 & 1.5 & 3.9 & 15.3 & 5.7 & 0.1 & 3.8 & 14.1 & 4.9 & 0.0 & 22.0 & 2.9 
\\
Whisper medium 
& 0.9 & 8.1 & 9.6 & 10.0 & 8.5 & 23.5 & 13.8 & 0.5 & 10.9 & 23.2 & 11.2 & 0.2 & 29.1 & 12.7 
\\
Whisper large 
& 1.2 & 9.3 & 12.0 & 12.5 & 9.4 & 26.4 & 16.5 & 1.0 & 13.1 & 25.5 & 12.8 & 0.5 & 30.5 & 12.9 
\\
Whisper large-v2 
& 1.0 & 11.0 & 14.0 & 14.3 & 10.2 & 27.7 & 16.7 & 1.0 & 12.9 & 27.3 & 13.5 & 0.4 & 31.4 & 16.1 
\\
\bottomrule
\end{tabular}
\vskip 0.2em
\begin{tabular}{l|rrrrrrrrrrrrrr}
\toprule
\setlength{\tabcolsep}{-5pt}\renewcommand{\arraystretch}{1.2}
Model
& \multicolumn{1}{c}{\rotatebox[origin=rc]{270}{Dutch}}
& \multicolumn{1}{c}{\rotatebox[origin=rc]{270}{Occitan}}
& \multicolumn{1}{c}{\rotatebox[origin=rc]{270}{Punjabi}}
& \multicolumn{1}{c}{\rotatebox[origin=rc]{270}{Polish}}
& \multicolumn{1}{c}{\rotatebox[origin=rc]{270}{Pashto}}
& \multicolumn{1}{c}{\rotatebox[origin=rc]{270}{Portuguese}}
& \multicolumn{1}{c}{\rotatebox[origin=rc]{270}{Romanian}}
& \multicolumn{1}{c}{\rotatebox[origin=rc]{270}{Russian}}
& \multicolumn{1}{c}{\rotatebox[origin=rc]{270}{Sindhi}}
& \multicolumn{1}{c}{\rotatebox[origin=rc]{270}{Slovak}}
& \multicolumn{1}{c}{\rotatebox[origin=rc]{270}{Slovenian}}
& \multicolumn{1}{c}{\rotatebox[origin=rc]{270}{Shona}}
& \multicolumn{1}{c}{\rotatebox[origin=rc]{270}{Somali}}
& \multicolumn{1}{c}{\rotatebox[origin=rc]{270}{Serbian}}
\\ \midrule
Whisper tiny 
& 2.7 & 1.7 & 0.3 & 0.8 & 0.3 & 12.1 & 1.0 & 3.1 & 0.5 & 0.7 & 0.3 & 0.1 & 0.0 & 0.6 
\\
Whisper base 
& 7.5 & 4.2 & 1.1 & 5.1 & 0.4 & 22.4 & 4.9 & 12.1 & 0.7 & 4.6 & 1.3 & 0.3 & 0.1 & 5.4 
\\
Whisper small 
& 15.9 & 9.5 & 4.4 & 14.0 & 0.8 & 31.2 & 18.3 & 19.7 & 2.0 & 14.4 & 6.9 & 0.6 & 0.1 & 19.3 
\\
Whisper medium 
& 21.6 & 15.9 & 12.8 & 19.0 & 2.1 & 35.9 & 26.6 & 24.8 & 5.5 & 22.7 & 14.0 & 1.4 & 0.4 & 27.7 
\\
Whisper large 
& 22.8 & 16.8 & 14.6 & 21.4 & 3.7 & 37.4 & 29.1 & 26.7 & 5.9 & 25.1 & 16.9 & 1.8 & 0.5 & 30.5 
\\
Whisper large-v2 
& 24.0 & 20.2 & 15.7 & 22.3 & 3.4 & 38.1 & 31.5 & 27.8 & 5.7 & 26.1 & 17.0 & 1.8 & 0.7 & 32.5 
\\
\bottomrule
\end{tabular}
\vskip 0.2em
\begin{tabular}{l|rrrrrrrrrrrr}
\toprule
\setlength{\tabcolsep}{-5pt}\renewcommand{\arraystretch}{1.2}
Model
& \multicolumn{1}{c}{\rotatebox[origin=rc]{270}{Swedish}}
& \multicolumn{1}{c}{\rotatebox[origin=rc]{270}{Swahili}}
& \multicolumn{1}{c}{\rotatebox[origin=rc]{270}{Tamil}}
& \multicolumn{1}{c}{\rotatebox[origin=rc]{270}{Telugu}}
& \multicolumn{1}{c}{\rotatebox[origin=rc]{270}{Tajik}}
& \multicolumn{1}{c}{\rotatebox[origin=rc]{270}{Thai}}
& \multicolumn{1}{c}{\rotatebox[origin=rc]{270}{Turkish}}
& \multicolumn{1}{c}{\rotatebox[origin=rc]{270}{Ukrainian}}
& \multicolumn{1}{c}{\rotatebox[origin=rc]{270}{Urdu}}
& \multicolumn{1}{c}{\rotatebox[origin=rc]{270}{Uzbek}}
& \multicolumn{1}{c}{\rotatebox[origin=rc]{270}{Vietnamese}}
& \multicolumn{1}{c}{\rotatebox[origin=rc]{270}{Yoruba}}
\\ \midrule
Whisper tiny 
& 1.8 & 0.1 & 0.2 & 0.3 & 0.2 & 0.2 & 0.2 & 1.2 & 0.4 & 0.0 & 0.1 & 0.2 
\\
Whisper base 
& 9.1 & 0.1 & 0.4 & 0.4 & 0.2 & 0.7 & 2.4 & 6.9 & 1.5 & 0.2 & 0.9 & 0.5 
\\
Whisper small 
& 22.9 & 0.1 & 2.1 & 4.0 & 4.4 & 5.8 & 15.7 & 18.7 & 8.8 & 0.5 & 8.5 & 0.5 
\\
Whisper medium 
& 32.1 & 3.1 & 7.0 & 10.8 & 11.4 & 12.8 & 22.9 & 25.8 & 14.9 & 3.8 & 16.6 & 0.9 
\\
Whisper large 
& 33.1 & 5.3 & 8.5 & 10.9 & 13.0 & 15.2 & 25.7 & 28.0 & 16.3 & 5.8 & 19.5 & 1.2 
\\
Whisper large-v2 
& 35.3 & 7.2 & 9.2 & 12.5 & 14.5 & 16.1 & 26.6 & 29.4 & 17.2 & 6.0 & 20.4 & 1.4 
\\
\bottomrule
\end{tabular}
\caption{BLEU scores on Fleurs}\label{tab:fleurs-translate}
\vspace{-2em}
\end{table}

\subsubsection{CoVOST 2}
\begin{table}[H]\centering
\scriptsize
\vskip 0.2em
\begin{tabular}{l|rrrrrrrrrrrrr}
\toprule
\setlength{\tabcolsep}{-5pt}\renewcommand{\arraystretch}{1.2}
Model
& \multicolumn{1}{c}{\rotatebox[origin=rc]{270}{Arabic}}
& \multicolumn{1}{c}{\rotatebox[origin=rc]{270}{Catalan}}
& \multicolumn{1}{c}{\rotatebox[origin=rc]{270}{Welsh}}
& \multicolumn{1}{c}{\rotatebox[origin=rc]{270}{German}}
& \multicolumn{1}{c}{\rotatebox[origin=rc]{270}{Spanish}}
& \multicolumn{1}{c}{\rotatebox[origin=rc]{270}{Estonian}}
& \multicolumn{1}{c}{\rotatebox[origin=rc]{270}{Persian}}
& \multicolumn{1}{c}{\rotatebox[origin=rc]{270}{French}}
& \multicolumn{1}{c}{\rotatebox[origin=rc]{270}{Indonesian}}
& \multicolumn{1}{c}{\rotatebox[origin=rc]{270}{Italian}}
& \multicolumn{1}{c}{\rotatebox[origin=rc]{270}{Japanese}}
& \multicolumn{1}{c}{\rotatebox[origin=rc]{270}{Latvian}}
& \multicolumn{1}{c}{\rotatebox[origin=rc]{270}{Mongolian}}
\\ \midrule
Whisper tiny 
& 0.2 & 4.9 & 0.4 & 4.0 & 10.5 & 0.2 & 0.1 & 6.1 & 0.3 & 5.1 & 0.3 & 0.1 & 0.1 
\\
Whisper base 
& 1.2 & 11.0 & 0.5 & 11.7 & 21.3 & 0.3 & 0.1 & 15.4 & 4.9 & 13.0 & 4.9 & 0.5 & 0.1 
\\
Whisper small 
& 17.7 & 22.3 & 1.0 & 25.3 & 33.0 & 2.4 & 4.9 & 27.3 & 27.6 & 24.0 & 17.3 & 1.4 & 0.2 
\\
Whisper medium 
& 30.6 & 29.2 & 12.1 & 33.2 & 38.4 & 11.4 & 15.5 & 33.6 & 42.3 & 29.5 & 24.6 & 9.7 & 0.2 
\\
Whisper large 
& 35.5 & 30.3 & 16.1 & 34.3 & 38.0 & 13.4 & 17.5 & 34.4 & 45.4 & 29.1 & 24.2 & 10.5 & 0.3 
\\
Whisper large-v2 
& 39.7 & 31.8 & 21.5 & 36.3 & 40.1 & 15.0 & 19.3 & 36.4 & 48.1 & 30.9 & 26.1 & 13.9 & 0.1 
\\
\bottomrule
\end{tabular}
\vskip 0.2em
\begin{tabular}{l|rrrrrrrr}
\toprule
\setlength{\tabcolsep}{-5pt}\renewcommand{\arraystretch}{1.2}
Model
& \multicolumn{1}{c}{\rotatebox[origin=rc]{270}{Dutch}}
& \multicolumn{1}{c}{\rotatebox[origin=rc]{270}{Portuguese}}
& \multicolumn{1}{c}{\rotatebox[origin=rc]{270}{Russian}}
& \multicolumn{1}{c}{\rotatebox[origin=rc]{270}{Slovenian}}
& \multicolumn{1}{c}{\rotatebox[origin=rc]{270}{Swedish}}
& \multicolumn{1}{c}{\rotatebox[origin=rc]{270}{Tamil}}
& \multicolumn{1}{c}{\rotatebox[origin=rc]{270}{Turkish}}
& \multicolumn{1}{c}{\rotatebox[origin=rc]{270}{Chinese}}
\\ \midrule
Whisper tiny 
& 4.3 & 9.5 & 5.7 & 0.4 & 2.0 & 0.1 & 0.2 & 0.4 
\\
Whisper base 
& 12.4 & 23.2 & 16.1 & 1.4 & 10.5 & 0.4 & 2.8 & 1.4 
\\
Whisper small 
& 28.1 & 40.6 & 30.9 & 9.2 & 29.9 & 1.7 & 16.8 & 6.8 
\\
Whisper medium 
& 38.1 & 48.7 & 39.4 & 17.7 & 39.5 & 2.9 & 27.0 & 14.0 
\\
Whisper large 
& 39.3 & 48.6 & 41.6 & 23.9 & 40.3 & 3.7 & 26.7 & 17.1 
\\
Whisper large-v2 
& 41.2 & 51.6 & 43.3 & 21.6 & 42.9 & 4.2 & 28.3 & 18.0 
\\
\bottomrule
\end{tabular}
\caption{BLEU scores on CoVoST2}\label{tab:covost2-translate}
\end{table}

\subsection{Long-form Transcription}
\begin{table}[H]\centering
\scriptsize
\begin{tabular}{l|rrrrrrr}
\toprule
\setlength{\tabcolsep}{2pt}\renewcommand{\arraystretch}{1.2}
Model
& \multicolumn{1}{c}{\rotatebox[origin=rc]{270}{TED-LIUM3}}
& \multicolumn{1}{c}{\rotatebox[origin=rc]{270}{Meanwhile}}
& \multicolumn{1}{c}{\rotatebox[origin=rc]{270}{Kincaid46}}
& \multicolumn{1}{c}{\rotatebox[origin=rc]{270}{Rev16}}
& \multicolumn{1}{c}{\rotatebox[origin=rc]{270}{Earnings-21}}
& \multicolumn{1}{c}{\rotatebox[origin=rc]{270}{Earnings-22}}
& \multicolumn{1}{c}{\rotatebox[origin=rc]{270}{CORAAL}}
\\ \midrule
Whisper tiny.en 
& 5.5 & 12.8 & 13.8 & 15.1 & 17.0 & 22.0 & 30.3 
\\
Whisper tiny 
& 6.8 & 15.5 & 16.7 & 17.0 & 18.7 & 24.4 & 33.1 
\\
Whisper base.en 
& 4.6 & 9.4 & 11.2 & 13.2 & 12.5 & 16.6 & 25.2 
\\
Whisper base 
& 4.8 & 12.2 & 12.2 & 14.5 & 13.5 & 18.4 & 26.9 
\\
Whisper small.en 
& 4.6 & 6.0 & 9.4 & 12.0 & 10.8 & 14.0 & 21.9 
\\
Whisper small 
& 4.2 & 6.9 & 10.1 & 12.1 & 11.1 & 14.3 & 22.3 
\\
Whisper medium.en 
& 3.6 & 5.2 & 8.9 & 11.9 & 10.2 & 13.3 & 20.6 
\\
Whisper medium 
& 3.8 & 5.4 & 8.6 & 11.4 & 10.3 & 13.2 & 20.3 
\\
Whisper large 
& 3.8 & 5.3 & 8.8 & 11.0 & 10.3 & 13.4 & 20.4 
\\
Whisper large-v2
& 3.5 & 5.1 & 8.8 & 11.3 & 9.7 & 12.6 & 19.6
\\
\midrule
wav2vec2-base-100h 
& 17.6 & 27.7 & 39.3 & 35.2 & 45.7 & 57.1 & 55.4 
\\
wav2vec2-base-960h 
& 12.8 & 19.7 & 32.9 & 29.8 & 37.3 & 46.8 & 49.1 
\\
wav2vec2-large-960h-lv60-self 
& 7.2 & 11.4 & 21.1 & 21.3 & 21.7 & 28.0 & 36.7 
\\
wav2vec2-large-960h 
& 10.1 & 16.4 & 27.4 & 26.4 & 30.4 & 40.1 & 43.5 
\\
wav2vec2-large-robust-ft-libri-960h 
& 8.8 & 15.2 & 22.9 & 23.4 & 23.0 & 31.0 & 36.8 
\\
hubert-large-ls960-ft 
& 8.1 & 12.9 & 22.4 & 23.4 & 23.0 & 30.6 & 37.9 
\\
hubert-xlarge-ls960-ft 
& 8.1 & 12.5 & 22.9 & 23.2 & 23.1 & 31.3 & 38.1 
\\
stt\_en\_conformer\_ctc\_large 
& 4.0 & 9.8 & 13.1 & 14.5 & 12.6 & 17.6 & 25.1 
\\
stt\_en\_conformer\_transducer\_xlarge 
& 5.3 & 10.6 & 17.1 & 19.8 & 16.2 & 19.7 & 38.9 
\\
\bottomrule
\end{tabular}
\caption{Long-form English transcription WER (\%)}\label{tab:longform-full-table}
\end{table}

\vfill
\pagebreak

\section{Training Dataset Statistics}

\begin{figure}[h!]
\begin{center}
\centerline{\includegraphics[width=1.0\textwidth]{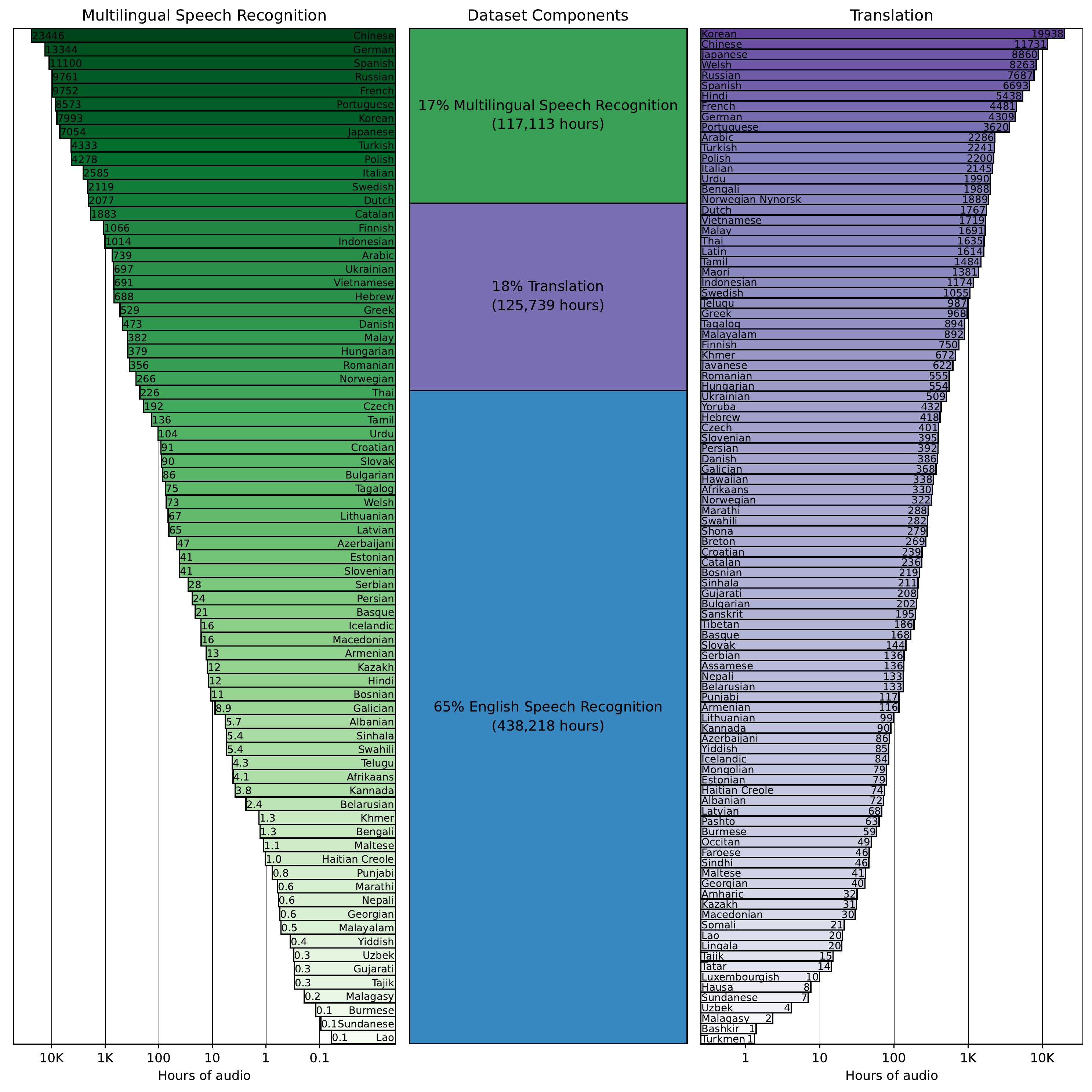}}
\caption{Training dataset statistics}
\label{dataset_stats}
\end{center}
\vspace{-1em}
\end{figure}

\vfill
\pagebreak

\section{Hyperparameters}\label{sec:hyperparameters}

\begin{table}[H]
\centering
\begin{tabular}{l|c} \toprule
    Hyperparameter & Value \\
    \midrule
    Updates & 1048576 \\
    Batch Size & 256 \\
    Warmup Updates & 2048 \\
    Max grad norm & 1.0 \\
    Optimizer & AdamW \\
    $\beta_{1}$ & 0.9 \\
    $\beta_{2}$ & 0.98 \\
    $\epsilon$ & $10^{-6}$ \\
    Weight Decay & 0.1 \\
    Weight Init & Gaussian Fan-In \\
    Learning Rate Schedule & Linear Decay \\
    Speechless audio subsample factor & $10\times$ \\
    Condition on prior text rate & 50\% \\
    \bottomrule
\end{tabular}
\caption{Whisper training hyperparameters.}\label{tab:hyperparams-training}
\end{table}

\begin{table}[H]
\centering
\begin{tabular}{l|c} \toprule
    Hyperparameter & Value \\
    \midrule
    Updates & 655360 \\
    Batch Size & 1024 \\
    BPE Dropout & 0.1 \\
    Stochastic Depth & 0.1 \\
    SpecAugment Policy & LibriSpeech Basic \\
    \bottomrule
\end{tabular}
\caption{Hyperparameters changed for Whisper Large V2.}\label{tab:hyperparams-training2}
\end{table}

\begin{table}[H]
\centering
\begin{tabular}{l|c} \toprule
    Model & Max Learning Rate \\
    \midrule
    Tiny & $1.5\times10^{-3}$\\
    Base & $1\times10^{-3}$\\
    Small & $5\times10^{-4}$\\
    Medium & $2.5\times10^{-4}$ \\
    Large & $1.75\times10^{-4}$ \\
    Large V2 & $2.0\times10^{-4}$ \\
    \bottomrule
\end{tabular}
\caption{Whisper model learning rates.}\label{tab:hyperparams-learning-rates}
\end{table}

\end{document}